\RequirePackage{fixltx2e}
\documentclass[11pt,a4paper]{article}
\pdfoutput=1
\usepackage[utf8]{inputenc}
\usepackage{jcappub}
\hypersetup{unicode=true,bookmarksopen=true}

\usepackage{mathtools}
\usepackage{lmodern}
\usepackage[T1]{fontenc}
\usepackage{bbm}
\DeclareMathAlphabet{\mathbfi}{OML}{cmm}{b}{it}

\let\originalleft\left
\let\originalright\right
\renewcommand{\left}{\mathopen{}\mathclose\bgroup\originalleft}
\renewcommand{\right}{\aftergroup\egroup\originalright}

\renewcommand{\vec}[1]{{\ifnum9<1#1\mathbf{#1}\else\ifcat\noexpand#1\relax\boldsymbol{#1}\else\mathbfi{#1}\fi\fi}}
\newcommand{\mathe}{\mathrm{e}}
\newcommand{\mathi}{\mathrm{i}}
\let\oldre\Re
\let\oldim\Im
\renewcommand{\Re}{\oldre\mathfrak{e}\,}
\renewcommand{\Im}{\oldim\mathfrak{m}\,}
\newcommand{\total}{\mathop{}\!\mathrm{d}}
\newcommand{\laplace}{\mathop{}\!\bigtriangleup}
\newcommand{\abs}[1]{{\left\lvert{#1}\right\rvert}}
\newcommand{\eqend}[1]{\,#1}
\newcommand{\bigo}[1]{\mathcal{O}\left({#1}\right)}
\newcommand{\Ein}{\operatorname{Ein}}
\DeclareMathOperator*{\distlim}{d-lim}

\allowdisplaybreaks

\begin{document}

\title{Quantum corrections for spinning particles in de~Sitter}

\author[1]{Markus B. Fröb}
\affiliation[1]{Department of Mathematics, University of York, Heslington, York, YO10 5DD, United Kingdom}

\author[2]{and Enric Verdaguer}
\affiliation[2]{Departament de Física Quàntica i Astrofísica, Institut de Ciències del Cosmos (ICC), Universitat de Barcelona (UB), C/ Martí i Franquès 1, 08028 Barcelona, Spain}

\emailAdd{mbf503@york.ac.uk}

\emailAdd{enric.verdaguer@ub.edu}

\abstract{We compute the one-loop quantum corrections to the gravitational potentials of a spinning point particle in a de~Sitter background, due to the vacuum polarisation induced by conformal fields in an effective field theory approach. We consider arbitrary conformal field theories, assuming only that the theory contains a large number $N$ of fields in order to separate their contribution from the one induced by virtual gravitons. The corrections are described in a gauge-invariant way, classifying the induced metric perturbations around the de~Sitter background according to their behaviour under transformations on equal-time hypersurfaces. There are six gauge-invariant modes: two scalar Bardeen potentials, one transverse vector and one transverse traceless tensor, of which one scalar and the vector couple to the spinning particle. The quantum corrections consist of three different parts: a generalisation of the flat-space correction, which is only significant at distances of the order of the Planck length; a constant correction depending on the undetermined parameters of the renormalised effective action; and a term which grows logarithmically with the distance from the particle. This last term is the most interesting, and when resummed gives a modified power law, enhancing the gravitational force at large distances. As a check on the accuracy of our calculation, we recover the linearised Kerr-de~Sitter metric in the classical limit and the flat-space quantum correction in the limit of vanishing Hubble constant.}


\maketitle

\section{Introduction}

Since all of the classical tests of Einstein's general relativity concern only small deviations from the Newtonian behaviour, and can thus be derived from particle motion in a corrected Newtonian potential, it seems fitting to also study a quantum-corrected Newtonian potential to derive effects due to a quantum theory of gravity. While a full theory of quantum gravity does not exist yet, and various competing approaches are being considered, one can nevertheless make predictions by treating quantum gravity (possibly coupled to other matter fields) as an effective quantum field theory~\cite{donoghue1994b,burgess2004}, able to describe quantum gravitational phenomena at energies well below some cutoff scale, which is essentially the Planck scale. In fact, quantum corrections to the Newton potential have been studied by many authors~\cite{radkowski1970,schwinger1968,duff1974,capperduffhalpern1974,capperduff1974,donoghue1994a,donoghue1994b,muzinichvokos1995,hamberliu1995,akhundovbelluccishiekh1997,dalvitmazzitelli1997,duffliu2000a,duffliu2000b,kirilinkhriplovich2002,khriplovichkirilin2003,bjerrumbohrdonoghueholstein2003a,bjerrumbohrdonoghueholstein2003b,satzmazzitellialvarez2005,holsteinross2008,parkwoodard2010,marunovicprokopec2011,marunovicprokopec2012,burnspilaftsis2015,bjerrumbohretal2016}, with the result that the effects are there, but too small to be observed at present. One can understand this conclusion from dimensional analysis alone: the expansion parameter for perturbative quantum gravity is the dimensionful Newton's constant $G_\text{N}$ (in the usual units where $\hbar = c = 1$, or $\hbar G_\text{N}/c^3$ otherwise), which is the square of the Planck length $\ell_\text{Pl}$. Since the only other scale in the problem is the distance $r$ from the source, the relative corrections must be given by a numerical constant (which is expected to be of order unity) times $G_\text{N}/r^2 = \ell_\text{Pl}^2/r^2$. Plugging in the numbers, one quickly realises that any correction is completely insignificant. One can slightly ameliorate the situation by considering gravity interacting with $N$ matter fields, where the above corrections get multiplied by $N$, and then take $N$ large. However, the standard model tells us that $N \approx 10^2$, which is not enough to overcome the smallness of Newton's constant.

The above picture changes considerably once quantum corrections are considered on a non-trivial background, different from flat space. In this case, the background spacetime furnishes another scale which can combine with Newton's constant to form a dimensionless quantity. A multitude of other functional forms of the corrections to the Newton potential are then possible, even constant corrections or ones which grow with the distance from the particle. Naturally, one has to perform concrete calculations to see if such corrections are actually present, and in case they are work out the numerics. Especially important in this context are quantum effects in de~Sitter spacetime, which is a very good approximation for most of the inflationary period in the standard cosmological model~\cite{mukhanov,kazanas1980,sato1981,guth1981,linde1982,albrechtsteinhardt1982}, and also models our present accelerated universe~\cite{perlmutteretal1998,knopetal2003,eisensteinetal2005,astieretal2006,kowalskietal2008}. For static point sources in de~Sitter spacetime, corrections to the Newton potential have been calculated very recently~\cite{wangwoodard2015b,parkprokopecwoodard2016,froebverdaguer2016a}, obtaining contributions which grow logarithmically with either time or distance, and are thus potentially much larger than in flat space.

The effective quantum gravity theory is described by a bare action for the metric and the matter fields, consisting of a series of diffeomorphism invariant scalars. The lowest-order terms are the familiar Einstein-Hilbert action with cosmological constant, while terms with more derivatives and/or powers of curvature come with additional powers of Newton's constant/the Planck length. The corresponding coupling parameters must be obtained from additional experiments, and since one needs more and more terms at each loop order, quantum gravity is perturbatively non-renormalisable. However, since they are suppressed relative to the lower-order terms, at low scales only few of them are needed, and one may obtain reliable predictions from the effective theory at those low scales. In some applications, the situation is even better: in general, at one-loop order one needs two additional counterterms quadratic in the curvature tensors (which may be taken to be the square of the Ricci scalar and the square of the Weyl tensor), and correspondingly one has two undetermined free parameters corresponding to the finite part of those counterterms. For certain observables in inflation, the contribution of those finite parts quickly red-shifts and is negligible at late times~\cite{miaowoodard2006,wangwoodard2015a}. Thus, completely unambiguous predictions can be obtained for those observables in the late-time limit.

In the recent work~\cite{froebverdaguer2016a}, we computed the leading quantum corrections to the gravitational potentials of a point mass in a de~Sitter background due to the coupling of the metric perturbations with conformal fields in an arbitrary conformal field theory. In addition to the Newton potential, which is obtained from the time-time component of the perturbed metric, there is another gauge-invariant variable that is classically constrained to be equal to the Newton potential, but receives quantitatively different quantum corrections. To isolate the contribution from conformal fields, in addition to working in the effective field theory approach we made a large-$N$ expansion, where $N$ is the number of conformal fields which we assumed to be large. The large-$N$ expansion then amounts, after rescaling Newton's constant, to a saddle point expansion of the path integral, in which graviton loops are suppressed by factors of $1/N$ relative to matter loops~\cite{tomboulis1977,hartlehorowitz1981,hurouraverdaguer2004}. The result shows that besides the quantum effects analogous to those found in flat spacetime, namely that the classical gravitational potential gets very small corrections proportional to $\ell_\text{Pl}^2/\hat{r}^2$, where $\hat{r}$ is the physical distance from the source, there are two new effects. The first is a constant shift which depends on the undetermined parameters of the renormalised effective action, and could be interpreted as an additional (finite but scale-dependent) renormalisation of Newton's constant. The second effect is a term that grows logarithmically with the distance, of the form $\ell_\text{Pl}^2 H^2 \ln \hat{r}$ with the Hubble parameter $H$, the new background scale in de Sitter spacetime. However, this logarithmic growth may be an artefact of perturbation theory, which one should only consider as valid up to distances $\hat{r}$ that keep this term bounded by one. Since only one-loop corrections were considered, we cannot make statements beyond one-loop perturbation theory. Nevertheless, to one-loop order this term leads to a modification of the $1/\hat{r}$ Newtonian gravitational law to a $1/\hat{r}^\alpha$ one where $\alpha = 1 - c \ell_\text{Pl}^2 H^2$ with $c > 0$, which means that the potential decays slower at large distances, indicating an enhancement of the gravitational attraction due to quantum effects.\footnote{\label{errnote}Note that due to a misprint, in the abstract and discussion of Ref.~\cite{froebverdaguer2016a} it is erroneously mentioned that the decay is faster. The error has been corrected in the arXiv version.} Moreover, the positivity of $c$ is guaranteed by unitarity~\cite{osbornpetkou1994}, such that this large-distance enhancement, even though small, is a universal effect.

The aim of this work is to extend the above results to spinning particles in de~Sitter spacetime, i.e., to calculate the quantum corrections to the gravitational potentials of a spinning point mass in de~Sitter due to loop corrections of conformal fields. By classifying the metric perturbations according to their behaviour under transformation on equal-time hypersurfaces into scalar, vector and tensor perturbations, one can construct six gauge-invariant variables: two scalar, one transverse vector and one transverse and traceless tensor potential. A spinless point particle only couples to one of the scalar potentials (the one that becomes the Newtonian potential in the non-relativistic limit), but a spinning particle also couples to the vector-type potential. This vector potential is responsible for the Lense-Thirring effect, which may also be interpreted as a long distance effect of the Kerr metric.

It seems that the first computation of quantum corrections to the gravitational potential of spinning particles was performed by Donoghue et al.~\cite{donoghueetal2002}, who found the long distance quantum corrections to the gravitational field of charged particles with and without spin. These authors work also in the framework of quantum gravity as an effective field theory, but since their approach is quite different from ours it is worth to explain their method in some detail. The first step in their approach is the calculation of the in-out matrix elements of the stress tensor describing the radiative corrections due to photons and gravitons on a static charged particle. Fourier transforming with respect to the difference $\vec{q}$ in spatial momenta between the in and the out state, they obtain an effective quantum stress tensor in position space, which in the next step is used as a source in the linearised classical Einstein equations. It turns out that the one-loop contributions from virtual massless photons lead to non-analytic terms in the matrix elements (of the form $\sqrt{\vec{q}^2}$ and $\ln \vec{q}^2$), which determine long-range corrections to the classical stress tensor of the charged particle. The metric perturbations thus obtained are the sum of three parts: a) the Newtonian potential, b) another classical term, which is the gravitational contribution associated to the electromagnetic field of the charged particle, and c) an extra term. The two classical ones reproduce the long-distance form of the Reissner-Nordstr{\"o}m and Kerr-Newman metrics, while the extra term is the quantum correction.

In contrast, our approach is based on the computation of the effective gravitational action for the gravitational field interacting with conformal fields, and deriving effective equations of motion for the metric perturbations from it. In contrast to the work of Donoghue et al.~\cite{donoghueetal2002}, who only study the cases of spin $0$ and $1/2$, we consider a classical spinning point mass with arbitrary spin. We furthermore work in the above-mentioned large-$N$ expansion, in which non-linear effects of the gravitational perturbations are suppressed by $1/N$, and we will thus not reproduce the classical non-linear terms in the long-distance behaviour of the Kerr metric. The reason for this choice is the notorious difficulty to define gauge-invariant local observables in (even perturbative) quantum gravity, once graviton loops are included. While in flat space one can reconstruct the scattering potential from the gauge-invariant S-Matrix (the inverse scattering method), this is not possible in de~Sitter spacetime where no S-Matrix exists~\cite{witten2001,bousso2005}. Even so, it took over 10 years from the first calculation~\cite{donoghue1994a} until~\cite{bjerrumbohrdonoghueholstein2003b,khriplovichkirilin2003} to find the right numerical coefficient for the correction due to graviton loops. Nevertheless, let us mention the recent works~\cite{miaowoodard2012,donnellygiddings2016,brunettietal2016}, where concrete proposals for gauge-invariant observables (up to an arbitrary fixed order in perturbation theory) were made. It would be illuminating to calculate the graviton corrections to the Newtonian potential in flat space using those observables, and to determine which one corresponds to the result obtained by the inverse scattering method. Those calculations could then be generalised to de~Sitter space and other curved spacetimes; however, this is vastly beyond the scope of our work.

The main advantage of the effective action approach is its applicability to non-trivial background spacetimes. In fact, since no S-Matrix exists in de~Sitter~\cite{witten2001,bousso2005}, the inverse scattering method cannot be applied, and solving the effective field equations is the only way to progress. The calculation of the effective action basically amounts to a one-loop computation of the graviton self-energy in the curved background. Adding the action for a point source, effective field equations for the metric perturbations can be obtained in the usual way by varying the effective action, and thus the gravitational response to a point source (or any other source, for that matter) can be studied~\cite{duff1974,duffliu2000a,satzmazzitellialvarez2005,parkwoodard2010,marunovicprokopec2011,marunovicprokopec2012}. As pointed out by Park and Woodard~\cite{parkwoodard2010} in this context, to obtain real and causal effective field equations it is necessary to compute the effective action using the Schwinger-Keldysh or ``in-in'' formalism~\cite{schwinger1961,keldysh1965,chousuhaoyu1985}. The underlying reason is the non-equivalence of the in- and out vacua due to particle production on time-dependent backgrounds (such as de~Sitter); in fact, the usual in-out formalism gives completely wrong results already in the trivial case of a free scalar field in de~Sitter, where part of the mass term is treated as a perturbation and then resummed to all orders~\cite{higuchilee2009}.

The paper is organised as follows: In section~\ref{sec_spinning} we give a brief review of the description of a test body with spin in de~Sitter spacetime, summarising the main results of the analysis of Obukhov and Puetzfeld~\cite{obukhovpuetzfeld2011} and providing explicit formulas for the Poincar{\'e} patch of de~Sitter, which we are taking as the background. In section~\ref{sec_effectiveaction}, we display the in-in effective action for the metric perturbations around this background (including the quantum corrections from conformal matter fields which were integrated out), based on the results of Campos and Verdaguer~\cite{camposverdaguer1994,camposverdaguer1996} for conformally coupled scalar fields which were subsequently generalised to general conformal matter~\cite{fprv2013,frv2014,froebverdaguer2016a}. The result only depends on gauge-invariant combinations (two scalar potentials, one vector and one tensor-like potential, according to their transformation properties on the background equal-time hypersurfaces), and their equations of motion are derived and subsequently solved in section~\ref{sec_eom}. In section~\ref{sec_comparison}, we compare the classical and the flat-space limit of our results to known works, and section~\ref{sec_discussion} presents the main result and conclusions. Some technical steps are delegated to the appendix. We use the ``+++'' convention of Ref.~\cite{mtw}, units such that $c = \hbar = 1$, and define $\kappa^2 \equiv 16 \pi G_\text{N}$ with Newton's constant $G_\text{N}$. Greek indices range over spacetime, while Latin ones are purely spatial.

\section{Action for spinning point particles}
\label{sec_spinning}

The first step in our analysis is to determine an action for a point particle with spin in a curved spacetime. It is known that in addition to the four-position $z^\mu(\tau)$ describing its world line, with $\tau$ an affine parameter, this involves an antisymmetric spin tensor $S^{\mu\nu}(\tau)$~\cite{mathisson1937,papapetrou1951,tulczyjew1959,tulczyjew1962,taub1964,dixon1964,trautmann1965}. Furthermore, the canonical momentum $p^\mu(\tau)$ can be different from the one in absence of spin (which would be given by $m u^\mu$ with $u^\mu \equiv \total z^\mu/\!\total \tau$), and the spin tensor can fulfil one of two constraints:
\begin{subequations}
\label{spin_condition}
\begin{align}
S^{\mu\nu} u_\mu &= 0 \eqend{,} \label{spin_condition_fp} \\
S^{\mu\nu} p_\mu &= 0 \eqend{.} \label{spin_condition_tu}
\end{align}
\end{subequations}
The first one is known as the Frenkel-Pirani condition~\cite{frenkel1926,pirani1956}, while the second one is the Tulczyjew condition~\cite{tulczyjew1959,tulczyjew1962}. In the literature~\cite{ohashi2003,steinhoff2010,blanchet2011}, one then finds an expression for the stress tensor of a spinning particle, which reads
\begin{equation}
\label{stress_tensor}
T^{\mu\nu}(x) = \int \delta(x-z(\tau)) p^{(\mu}(\tau) u^{\nu)}(\tau) \total \tau - \nabla_\alpha \int \delta(x-z(\tau)) S^{\alpha(\mu}(\tau) u^{\nu)}(\tau) \total \tau
\end{equation}
with the covariant $\delta$ distribution
\begin{equation}
\delta(x-y) \equiv \frac{\delta^n(x-y)}{\sqrt{-g(x)}} \eqend{.}
\end{equation}
From its covariant conservation, using that $\total/\!\total \tau = u^\mu \nabla_\mu$ we find the equation of motion for the particle (usually known as the Mathisson-Papapetrou equation~\cite{mathisson1937,papapetrou1951}), which reads
\begin{equation}
\label{pp_eom}
\frac{\total p_\alpha}{\total \tau} = - \frac{1}{2} R_{\alpha\beta\mu\nu} u^\beta S^{\mu\nu} \eqend{,}
\end{equation}
and the spin precession equation
\begin{equation}
\label{pp_spin}
\frac{\total S^{\mu\nu}}{\total \tau} = p^\mu u^\nu - p^\nu u^\mu \eqend{.}
\end{equation}
For a background de~Sitter spacetime, the Riemann tensor is given by $R_{\alpha\beta\mu\nu} = 2 H^2 g_{\alpha[\mu} g_{\nu]\beta}$, and thus the equation of motion reduces to
\begin{equation}
\label{pp_eom_ds}
\frac{\total p^\mu}{\total \tau} = - H^2 S^{\mu\nu} u_\nu \eqend{.}
\end{equation}

These equations have been recently solved by Obukhov and Puetzfeld~\cite{obukhovpuetzfeld2011}, and we summarise some of their results now. First note that by contracting equation~\eqref{pp_spin} with $u_\mu$, we obtain
\begin{equation}
\label{p_in_u_dsdt}
u_\mu \frac{\total S^{\mu\nu}}{\total \tau} = - m u^\nu + p^\nu
\end{equation}
with the mass parameter
\begin{equation}
\label{mass_param_fp}
m \equiv - p^\mu u_\mu \eqend{,}
\end{equation}
while contracting equation~\eqref{pp_eom} or~\eqref{pp_eom_ds} with $u_\alpha$ we get
\begin{equation}
\label{u_dp_dt_vanishing}
u_\alpha \frac{\total p^\alpha}{\total \tau} = 0 \eqend{.}
\end{equation}
Thus, for the time dependence of the mass parameter $m$ we calculate
\begin{equation}
\frac{\total m}{\total \tau} = - p^\mu \frac{\total u_\mu}{\total \tau} = - \left( u_\nu \frac{\total S^{\nu\mu}}{\total \tau} + m u^\mu \right) \frac{\total u_\mu}{\total \tau} = \frac{\total u_\mu}{\total \tau} \frac{\total}{\total \tau} \left( S^{\mu\nu} u_\nu \right) \eqend{,}
\end{equation}
such that $m$ is conserved if we impose the Frenkel-Pirani condition~\eqref{spin_condition_fp}. On the other hand, defining the mass parameter $M$ by
\begin{equation}
\label{mass_param_tu}
M^2 \equiv - p^\mu p_\mu \eqend{,}
\end{equation}
we obtain using the equation of motion~\eqref{pp_eom_ds}
\begin{equation}
\frac{\total M^2}{\total \tau} = - 2 p_\mu \frac{\total p^\mu}{\total \tau} = 2 H^2 p_\mu u_\nu S^{\mu\nu} \eqend{,}
\end{equation}
which is conserved for either the Frenkel-Pirani~\eqref{spin_condition_fp} or the Tulczyjew condition~\eqref{spin_condition_tu}. Also the magnitude of the spin tensor is conserved, as shown by
\begin{equation}
\frac{\total \left( S^{\mu\nu} S_{\mu\nu} \right)}{\total \tau} = 4 S^{\mu\nu} p_\mu u_\nu = - \frac{4}{H^2} p_\mu \frac{\total p^\mu}{\total \tau} = \frac{2}{H^2} \frac{\total M^2}{\total \tau} \eqend{.}
\end{equation}
Note that for vanishing spin $S^{\mu\nu} = 0$, one sees that $p^\mu = m u^\mu$ with constant $m$ is a solution of the spin precession equation~\eqref{pp_spin}, while the equation of motion~\eqref{pp_eom} reduces to the usual equation of motion $\total u^\mu/\!\total \tau = 0$ for a spinless point particle in the absence of external forces. Furthermore, in this case both mass parameters agree: $m = M$.

In general, the quantity
\begin{equation}
\label{conserved_xi_general}
\Xi \equiv \xi_\mu p^\mu + \frac{1}{2} S^{\mu\nu} \nabla_\mu \xi_\nu
\end{equation}
is conserved on solutions of the equations~\eqref{pp_eom} and~\eqref{pp_spin} for any Killing vector $\xi^\mu$, which can be verified straightforwardly using the Bianchi identities for the Riemann tensor and the fact that for the second derivative of any Killing vector we have~\cite{mtw}
\begin{equation}
\nabla_\alpha \nabla_\beta \xi_\mu = R_{\mu\beta\alpha\nu} \xi^\nu = 2 H^2 g_{\alpha[\mu} \xi_{\beta]} \eqend{.}
\end{equation}
This can be used to give a complete solution to the equation of motion~\eqref{pp_eom_ds} and the spin precession equation~\eqref{pp_spin}. We work in the conformally flat coordinate system of the cosmological or Poincar{\'e} patch of de~Sitter spacetime with the $n$-dimensional metric
\begin{equation}
\label{ds_metric}
\total s^2 = a^2(\eta) \left( - \total \eta^2 + \total \vec{x}^2 \right)
\end{equation}
and the scale factor $a(\eta) = (-H\eta)^{-1}$, where $H$ is the constant Hubble parameter. The Christoffel symbols read
\begin{equation}
\Gamma^\alpha_{\beta\gamma} = \left( \delta^\alpha_\gamma \delta^0_\beta + \delta^0_\gamma \delta^\alpha_\beta + \delta_0^\alpha \eta_{\beta\gamma} \right) H a \eqend{,}
\end{equation}
and we obtain the coordinate expressions for the Killing vectors by explicitly solving their defining equation $0 = \nabla_{(\mu} \xi_{\nu)} = \partial_{(\mu} \xi_{\nu)} - \Gamma^\alpha_{\mu\nu} \xi_\alpha$. There are $(n-1)+(n-1)(n-2)/2$ Killing vectors corresponding to spatial translations and rotations, 1 which in the flat-space limit reduces to a time translation and $(n-1)$ generalised boosts, which together comprise the maximum number of $n (n+1)/2$ Killing vectors an $n$-dimensional spacetime can have. Their components are given by
\begin{subequations}
\label{killing_vectors}
\begin{align}
\text{(spatial translations)}&\ &\xi^\mu_{(\text{st},i)} &= \delta_i^\mu \eqend{,} \\
\text{(spatial rotations)}&\ &\xi^\mu_{(\text{sr},ij)} &= 2 x^k \eta_{k[i} \delta_{j]}^\mu \eqend{,} \\
\text{(gen. time translation)}&\ &\xi^\mu_{(\text{tt})} &= a^{-1} \delta_0^\mu - H x^i \delta_i^\mu \eqend{,} \\
\text{(gen. boosts)}&\ &\xi^\mu_{(\text{gb},i)} &= a^{-1} x^k \eta_{ki} \delta_0^\mu - H x^k \eta_{ki} x^j \delta_j^\mu + \frac{1}{2} \left[ H^{-1} ( 1 - a^{-2} ) + H \vec{x}^2 \right] \delta_i^\mu \eqend{,}
\end{align}
\end{subequations}
and plugging those expressions into equation~\eqref{conserved_xi_general}, we obtain the conserved quantities
\begin{subequations}
\label{conserved_xi}
\begin{align}
\Xi^{(\text{st})}_i &\equiv p_i + H a S^0{}_i \eqend{,} \\
\Xi^{(\text{sr})}_{ij} &\equiv 2 x^k \eta_{k[i} \left( p_{j]} + H a S^0{}_{j]} \right) + a^{-2} S_{ij} \eqend{,} \\
\Xi^{(\text{tt})} &\equiv a^{-1} p_0 - H x^i \left( p_i + H a S^0{}_i \right) \eqend{,} \\
\begin{split}
\Xi^{(\text{gb})}_i &\equiv a^{-1} x^k \eta_{ki} p_0 - H x^k \eta_{ki} x^j \left( p_j + H a S^0{}_j \right) \\
&\quad+ \frac{1}{2} \left[ H^{-1} ( 1 - a^{-2} ) + H \vec{x}^2 \right] \left( p_i + H a S^0{}_i \right) + a^{-1} S^0{}_i - H a^{-2} x^j S_{ij} \eqend{.}
\end{split}
\end{align}
\end{subequations}
These equations can be inverted to obtain expressions for the momenta $p_\mu$ and the spin tensor $S_{\mu\nu}$, and we obtain the explicit solution
\begin{subequations}
\label{sol_from_conserved}
\begin{align}
p_0 &= a \Xi^{(\text{tt})} + H a x^i \Xi^{(\text{st})}_i \eqend{,} \\
p_i &= \Xi^{(\text{st})}_i - H a S^0{}_i \eqend{,} \\
\begin{split}
S^0{}_i &= a \Xi^{(\text{gb})}_i - \frac{a}{2} \left[ H^{-1} ( 1 - a^{-2} ) - H \vec{x}^2 \right] \Xi^{(\text{st})}_i - a \eta_{ik} x^k \left[ \Xi^{(\text{tt})} + H x^j \Xi^{(\text{st})}_j \right] + H a x^j \Xi^{(\text{sr})}_{ij} \eqend{,}
\end{split} \\
S_{ij} &= a^2 \Xi^{(\text{sr})}_{ij} - 2 a^2 x^k \eta_{k[i} \Xi^{(\text{st})}_{j]} \eqend{.}
\end{align}
\end{subequations}

We are especially interested in solutions for a particle at rest at the origin, where $x^i(\tau) = 0$, and thus $u^i(\tau) = 0$. From the normalisation $u^\mu u_\mu = -1$, which fixes $\tau$ to be the proper time of the particle, we then obtain
\begin{equation}
u^0(\tau) = a^{-1}(\eta(\tau)) = \frac{\total \eta(\tau)}{\total \tau} \eqend{,}
\end{equation}
and thus
\begin{equation}
\eta(\tau) = - \frac{1}{H} \exp\left( - H \tau \right) \eqend{,} \quad a(\tau) = \exp\left( H \tau \right) \eqend{.}
\end{equation}
In this case, the above system~\eqref{sol_from_conserved} reduces to
\begin{subequations}
\label{sol_origin}
\begin{align}
p_0 &= a \Xi^{(\text{tt})} \eqend{,} \\
p_i &= \Xi^{(\text{st})}_i - H a S^0{}_i \eqend{,} \\
S^0{}_i &= a \Xi^{(\text{gb})}_i - \frac{a}{2 H} ( 1 - a^{-2} ) \Xi^{(\text{st})}_i \eqend{,} \\
S_{ij} &= a^2 \Xi^{(\text{sr})}_{ij} \eqend{.}
\end{align}
\end{subequations}
The derivatives of this explicit solution are easily calculated (remembering that $\total/\!\total \tau \equiv u^\alpha \nabla_\alpha$), and read
\begin{subequations}
\label{sol_origin_derivs}
\begin{align}
\frac{\total p_0}{\total \tau} &= 0 \eqend{,} \\
\frac{\total p_i}{\total \tau} &= - H^2 a S^0{}_i \eqend{,} \\
\frac{\total S^0{}_i}{\total \tau} &= H S^0{}_i - a^{-1} \Xi^{(\text{st})}_i \eqend{,} \\
\frac{\total S_{ij}}{\total \tau} &= 0 \eqend{,}
\end{align}
\end{subequations}
and comparing with the equation of motion~\eqref{pp_eom_ds} and the spin precession equation~\eqref{pp_spin} shows that both are fulfilled. Furthermore, one easily checks that both the Frenkel-Pirani condition~\eqref{spin_condition_fp} and the Tulczyjew condition~\eqref{spin_condition_tu} are satisfied for $\Xi^{(\text{gb})}_i = \Xi^{(\text{st})}_i = 0$. For the mass parameters~\eqref{mass_param_fp} and~\eqref{mass_param_tu} we then obtain
\begin{equation}
m = M = - \Xi^{(\text{tt})} \eqend{,}
\end{equation}
such that the full solution reads
\begin{subequations}
\begin{align}
p^0 &= m u^0 = m a^{-1} \eqend{,} \\
p^i &= u^i = S^0{}_i = 0 \eqend{,} \\
S_{ij} &= a^2 \Xi^{(\text{sr})}_{ij} \label{spin_tensor_sol}
\end{align}
\end{subequations}
with an arbitrary constant tensor $\Xi^{(\text{sr})}_{ij}$, which is by construction antisymmetric in the index pair $ij$ and which determines the spin of the particle as shown by equation~\eqref{spin_tensor_sol}. Evaluated on this solution, the stress tensor~\eqref{stress_tensor} reads
\begin{equation}
\begin{split}
T^{\mu\nu}(x) &= m \int \delta^n(x-z(\tau)) a^{-n-2} \delta^\mu_0 \delta^\nu_0 \total \tau - \nabla_\alpha \int \delta^n(x-z(\tau)) a^{-n-3} \eta^{i\alpha} \eta^{j(\mu} \delta^{\nu)}_0 \Xi^{(\text{sr})}_{ij} \total \tau \\
&= m a^{-n-1} \delta^\mu_0 \delta^\nu_0 \delta^{n-1}(\vec{x}) + a^{-n-2} \Xi^{(\text{sr})}_{ij} \delta_0^{(\mu} \eta^{\nu)[i} \eta^{j]k} \partial_k \delta^{n-1}(\vec{x}) \eqend{.}
\end{split} \raisetag{\baselineskip}
\end{equation}
The linearised action is then given as usual by introducing metric perturbations $h_{ab}$ via
\begin{equation}
\label{ds_metric_pert}
\tilde{g}_{\mu\nu} = a^2 g_{\mu\nu} = a^2 ( \eta_{\mu\nu} + h_{\mu\nu} ) \eqend{,}
\end{equation}
and coupling the stress tensor to the metric perturbation according to
\begin{equation}
\label{action_pp}
S_\text{PP} \equiv \frac{1}{2} \int a^{n+2} h_{\mu\nu} T^{\mu\nu} \total^n x \eqend{,}
\end{equation}
where the explicit factor of $a^{n+2}$ comes in because of the rescaling of the metric perturbation~\eqref{ds_metric_pert}, and the fact that for the background de~Sitter metric~\eqref{ds_metric} we have $\sqrt{-g} = a^n$.

Let us finally note that it is also possible to introduce a non-minimal spin--curvature coupling~\cite{deriglazovramirez2015a,deriglazovramirez2015b}, similar to the non-minimal coupling $\xi R \phi^2$ of a scalar field $\phi$, which would change the stress tensor~\eqref{stress_tensor}. However, in our case these corrections are of quadratic (or higher) order in the spin, i.e., they would induce changes involving a product of two (or more) tensors $\Xi^{(\text{sr})}_{ij}$. Since corrections quadratic in spin also would arise from the inclusion of graviton loops and graviton interaction vertices, which are suppressed by our use of the $1/N$-expansion, it does not seem sensible to keep terms coming from a non-minimal spin--curvature coupling either, and we leave a more detailed analysis to future work.

\section{The effective action}
\label{sec_effectiveaction}

It is well known that the diffeomorphism invariance of a gravitational theory with the metric $\tilde{g}_{\mu\nu}$ is, when expanding in perturbations around a background metric, equivalent to a gauge symmetry for the metric perturbations. For the background de~Sitter metric~\eqref{ds_metric} with perturbations~\eqref{ds_metric_pert}, this gauge symmetry takes the explicit form
\begin{equation}
\label{hmunu_gaugetrafo}
h_{\mu\nu} \to h_{\mu\nu} + 2 \partial_{(\mu} \xi_{\nu)} - 2 H a \eta_{\mu\nu} \xi_0 \eqend{,}
\end{equation}
where $\xi_\mu$ is a vector parametrising the gauge transformation. Since both the Einstein-Hilbert gravitational action as well as the usual matter actions are diffeomeorphism-invariant, their perturbative expansions, and consequently the effective action which is obtained after integrating out the matter fields, must be invariant under the transformation~\eqref{hmunu_gaugetrafo}, at least to lowest non-trivial order. In fact, it has been shown~\cite{abramobrandenbergermukhanov1997,nakamura2007,froebverdaguer2016a} that $h_{\mu\nu}$ can be split into a gauge-invariant part $h^\text{inv}_{\mu\nu}$ and a Lie derivative of a vector $X_\mu$, in the form
\begin{equation}
\begin{split}
\label{h_munu_inv_gauge}
h_{\mu\nu} &= h^\text{inv}_{\mu\nu} + a^{-2} \mathcal{L}_{a^2 X} \left( a^2 \eta_{\mu\nu} \right) \\
&= h^\text{inv}_{\mu\nu} + 2 \partial_{(\mu} X_{\nu)} - 2 H a \eta_{\mu\nu} X_0 \eqend{.}
\end{split}
\end{equation}
The gauge-invariant part is given by
\begin{equation}
\label{h_munu_inv_def}
h^\text{inv}_{\mu\nu} \equiv 2 \delta^0_\mu \delta^0_\nu \left( \Phi_\text{A} + \Phi_\text{H} \right) + 2 \eta_{\mu\nu} \Phi_\text{H} + 2 \delta^0_{(\mu} V_{\nu)} + h^\text{TT}_{\mu\nu} \eqend{,}
\end{equation}
where $\Phi_\text{A}$ and $\Phi_\text{H}$ are the two Bardeen potentials~\cite{bardeen1980}, while $V_\mu$ with $V_0 = 0$ is a spatial transverse vector, $\eta^{\mu\nu} \partial_\mu V_\nu = 0$ and $h^\text{TT}_{\mu\nu}$ with $h^\text{TT}_{0\mu} = 0$ is a spatial traceless, transverse tensor, $\eta^{\mu\nu} \partial_\mu h^\text{TT}_{\nu\rho} = 0 = \eta^{\mu\nu} h^\text{TT}_{\mu\nu}$. The gauge transformation~\eqref{hmunu_gaugetrafo} exclusively affects the vector $X_\mu$, namely under a gauge transformation we have $X_\mu \to X_\mu + \xi_\mu$. It follows that the effective action is invariant under the transformation~\eqref{hmunu_gaugetrafo} if it only depends on the invariant part $h^\text{inv}_{\mu\nu}$~\eqref{h_munu_inv_def}. The advantage of working with gauge-invariant variables from the start is a very practical one, since fewer equations must be solved. Moreover, to the order that we are working there is no mixing between scalar, vector and tensor perturbations such that we can treat each of those separately.

Using the decompositions~\eqref{h_munu_inv_gauge} and~\eqref{h_munu_inv_def} in the spinning particle action~\eqref{action_pp} and integrating by parts, we obtain in $n = 4$ dimensions
\begin{equation}
\label{action_pp_decomposed}
S_\text{PP} = \frac{1}{2} \int h^\text{inv}_{\mu\nu} T^{\mu\nu} a^6 \total^4 x = \int m a \Phi_\text{A} \delta^3(\vec{x}) \total^4 x + \frac{1}{2} \int \Xi^{(\text{sr})}_{ij} V^i \partial^j \delta^3(\vec{x}) \total^4 x \eqend{,}
\end{equation}
since covariant conservation of the stress tensor~\eqref{stress_tensor} (with respect to the de~Sitter background) ensures that the coupling~\eqref{action_pp} to the metric perturbations is gauge invariant, and thus does not depend on $X_\mu$. We see that the part that couples to scalar perturbations is unchanged from the case of a particle without spin~\cite{froebverdaguer2016a}, but the non-zero spin introduces a new coupling to vector perturbations. Since to the order we are working scalar and vector perturbations do not mix, we can focus on the vector perturbation and simply copy the final result for the scalar perturbations. There is no coupling to tensor perturbations, such that their equations of motion are the same as in the source-free case. Since the corresponding analysis was already performed in the work~\cite{fprv2013}, we from now on also ignore the tensor perturbations.

The effective action for metric perturbations interacting with massless, conformally coupled scalars in a FLRW background was first calculated by Campos and Verdaguer~\cite{camposverdaguer1994,camposverdaguer1996}, and later on generalised to general conformal matter~\cite{fprv2013,frv2014,froebverdaguer2016a}. As explained in the introduction, the in-in formalism~\cite{schwinger1961,keldysh1965,chousuhaoyu1985} has to be used to produce real and causal effective equations of motion, and a large-$N$ expansion has to be employed to separate effects due to matter loops from the effects of graviton loops~\cite{tomboulis1977,hartlehorowitz1981,hurouraverdaguer2004}. We refer the reader to the aforementioned works for details of the calculation, and only present here the end result for the effective action.

Starting with a bare Einstein-Hilbert action for gravity, the action for $N$ conformal matter fields, and a counterterm action to renormalise the ultraviolet divergences, one functionally integrates out the matter degrees of freedom to obtain the effective action for the metric perturbations, and then adds the point-particle action to describe their interaction with the spinning point particle. In the in-in formalism, the renormalised effective action
\begin{equation}
\label{seff}
S_\text{eff}\left[ \tilde{g}^\pm \right] = S_\text{loc,ren}\left[ a, h^+ \right] - S_\text{loc,ren}\left[ a, h^- \right] + \Sigma_\text{ren}\left[ h^\pm \right]
\end{equation}
depends on two types of fields, the ``$+$'' and the ``$-$'' ones, and the effective equations of motion are obtained by taking a variational derivative with respect to the ``$+$'' fields, and setting both types of fields equal to each other afterwards (see Ref.~\cite{froebverdaguer2016a} for more details). The first, local part of the effective action $S_\text{loc,ren}$ is given by
\begin{equation}
\label{slocren_def}
\begin{split}
S_\text{loc,ren} &\equiv \frac{1}{\kappa^2} \int \left( a^2 R - 6 a \nabla^\mu \nabla_\mu a - 2 \Lambda a^4 \right) \sqrt{-g} \total^4 x + \int \left( b C^2 + b' \mathcal{E}_4 \right) \ln a \sqrt{-g} \total^4 x \\
&\quad+ \frac{\beta}{12} \int \left( R - 6 a^{-1} \nabla^\mu \nabla_\mu a \right)^2 \sqrt{-g} \total^4 x - 4 b' \int G^{\mu\nu} a^{-2} ( \nabla_\mu a ) ( \nabla_\nu a ) \sqrt{-g} \total^4 x \\
&\quad+ 2 b' \int a^{-4} \left[ \nabla^\mu a \nabla_\mu a - 2 a \nabla^\mu \nabla_\mu a \right] \nabla^\nu a \nabla_\nu a \sqrt{-g} \total^4 x + S_\text{PP} \eqend{,}
\end{split}
\end{equation}
where the curvature tensors and covariant derivatives are calculated using the conformally related perturbed metric $g_{\mu\nu} = \eta_{\mu\nu} + h_{\mu\nu}$~\eqref{ds_metric_pert}, $C^2 \equiv C^{\mu\nu\rho\sigma} C_{\mu\nu\rho\sigma}$ is the square of the Weyl tensor $C_{\mu\nu\rho\sigma}$ and $\mathcal{E}_4 \equiv R^{\mu\nu\rho\sigma} R_{\mu\nu\rho\sigma} - 4 R^{\mu\nu} R_{\mu\nu} + R^2$ is the Euler density. The second, non-local part $\Sigma_\text{ren}$ reads
\begin{equation}
\label{sigmaren_def}
\begin{split}
\Sigma_\text{ren} &\equiv 2 b \iint C^+_{\mu\nu\rho\sigma}(x) C^{-\mu\nu\rho\sigma}(y) K(x-y) \total^4 x \total^4 y \\
&\quad+ b \iint C^+_{\mu\nu\rho\sigma}(x) C^{+\mu\nu\rho\sigma}(y) K^+(x-y; \bar{\mu}) \total^4 x \total^4 y \\
&\quad- b \iint C^-_{\mu\nu\rho\sigma}(x) C^{-\mu\nu\rho\sigma}(y) K^-(x-y; \bar{\mu}) \total^4 x \total^4 y \eqend{.}
\end{split}
\end{equation}
While the constant $\beta$ is arbitrary and must be determined by experiment, the parameters $b$ and $b'$ are the coefficients appearing in the trace anomaly in front of the square of the Weyl tensor and the Euler density. Both coefficients depend on the conformal theory under consideration; for $N_0$ free massless, conformally coupled scalar fields, $N_{1/2}$ free, massless Dirac spinor fields and $N_1$ free vector fields we have~\cite{duff1977}
\begin{subequations}
\label{coeffs_b}
\begin{align}
b &= \frac{N_0 + 6 N_{1/2} + 12 N_1}{1920 \pi^2} \eqend{,} \\
b' &= - \frac{N_0 + 11 N_{1/2} + 62 N_1}{5760 \pi^2} \eqend{.}
\end{align}
\end{subequations}
The non-local part $\Sigma_\text{ren}$ depends on non-local kernels $K$, which are given by their Fourier transforms
\begin{subequations}
\label{kernels_def}
\begin{align}
K(x) &\equiv - \mathi \pi \int \Theta(-p^2) \Theta(-p^0) \mathe^{\mathi p x} \frac{\total^4 p}{(2\pi)^4} \eqend{,} \\
K^\pm(x; \bar{\mu}) &\equiv \frac{1}{2} \int \left[ - \ln \abs{\frac{p^2}{\bar{\mu}^2}} \pm \mathi \pi \Theta(-p^2) \right] \mathe^{\mathi p x} \frac{\total^4 p}{(2\pi)^4} \eqend{.}
\end{align}
\end{subequations}
This result is valid in the MS renormalization scheme, where the renormalization scale $\bar{\mu}$ is chosen such that there is no term proportional to $C^2$ in the local part of the renormalized effective action $S_\text{loc,ren}$~\eqref{slocren_def} (except for the term involving $\ln a$ coming from the conformal transformation). However, the effective action is invariant under the renormalization group~\cite{toms1983} and cannot depend on the renormalization scale $\mu$. Thus, for all values of $\mu \neq \bar{\mu}$, an additional term appears in $S_\text{loc,ren}$, of the form $c(\mu) \int C^2 \total^4 x$ with $c(\mu) = - b \ln (\mu/\bar{\mu})$~\cite{froebverdaguer2016a}. While we will employ $\bar{\mu}$ in the following to shorten the formulas, we will restore the finite coefficient $c$ in the final results, i.e., perform the replacement
\begin{equation}
\label{barmu_replace}
b \ln \bar{\mu} \to b \ln \mu + c(\mu) \eqend{.}
\end{equation}

Setting the metric perturbation to zero after taking the variational derivatives gives the background equations of motion, which in our case determine the relation between the cosmological constant $\Lambda$ and the Hubble parameter $H$~\cite{froebverdaguer2016a},
\begin{equation}
\label{bg_eom_h}
\left. \frac{\delta S_\text{eff}[ a, h^\pm ]}{\delta h_{\mu\nu}^+} \right\rvert_{h^\pm = 0} = 0 \quad\Rightarrow\quad \Lambda = 3 H^2 \left( 1 + b' \kappa^2 H^2 \right) \eqend{,}
\end{equation}
As explained above, in this calculation we only focus on the vector perturbations. The effective action then consists of two parts~\cite{frv2014,froebverdaguer2016a}: The first one is local, and after inserting the decompositions~\eqref{h_munu_inv_gauge} and~\eqref{h_munu_inv_def} it reads [referring to the vector part with a superscript $(\text{V})$]
\begin{equation}
\label{seff_loc}
S^{(\text{V})}_\text{loc,ren} = S^{(\text{V})}_\text{PP} - \frac{1}{2 \kappa^2} \left[ 1 + (b+2b'-2\beta) \kappa^2 H^2 \right] \int a^2 V^k \laplace V_k \total^4 x - b \int \ln a \laplace V^k \partial^2 V_k \total^4 x \eqend{,}
\end{equation}
where $S^{(\text{V})}_\text{PP}$ is the vector part of the point particle action~\eqref{action_pp_decomposed}, $\laplace \equiv \delta^{ij} \partial_i \partial_j$, and $\partial^2 \equiv \eta^{\mu\nu} \partial_\mu \partial_\nu = - \partial_\eta^2 + \laplace$. The second part is non-local, and given by
\begin{equation}
\label{sigma}
\begin{split}
\Sigma^\text{(V)}_\text{ren} &= - b \iint \delta^{ij} \left[ \partial^2 V^+_i(x) \laplace V^-_j(y) + \laplace V^+_i(x) \partial^2 V^-_j(y) \right] K(x-y) \total^4 x \total^4 y \\
&\quad- \frac{b}{2} \iint \delta^{ij} \left[ \partial^2 V^+_i(x) \laplace V^+_j(y) + \laplace V^+_i(x) \partial^2 V^+_j(y) \right] K^+(x-y; \bar{\mu}) \total^4 x \total^4 y \\
&\quad+ \frac{b}{2} \iint \delta^{ij} \left[ \partial^2 V^-_i(x) \laplace V^-_j(y) + \laplace V^-_i(x) \partial^2 V^-_j(y) \right] K^-(x-y; \bar{\mu}) \total^4 x \total^4 y \eqend{,}
\end{split}
\end{equation}
with the kernels $K$ defined in equation~\eqref{kernels_def}. In contrast to the scalar case, the gauge-invariant vector perturbation $V_i$ is equal to the vector $v_i^\text{T}$ used in Ref.~\cite{frv2014} such that we could directly copy the above expressions, only generalising from the massless, conformally coupled scalar to a general conformal field theory as explained in our previous work~\cite{froebverdaguer2016a}.

\section{Effective equations and solutions for the gauge-invariant perturbations}
\label{sec_eom}

\subsection{Effective equations of motion}

The effective equations of motion for the gauge-invariant vector perturbation are now obtained by taking a variational derivative with respect to $V_k^+$ and setting $V_k^+ = V_k^- = V_k$ afterwards
\begin{equation}
\left. \frac{\delta S_\text{eff}^\text{(V)}[ a, h^\pm ]}{\delta V_k^+} \right\rvert_{V_k^+ = V_k^- = V_k} = 0 \eqend{.}
\end{equation}
Using that the kernel $K^+$ is symmetric, $K^+(x-y; \bar{\mu}) = K^+(y-x; \bar{\mu})$, this gives
\begin{equation}
\label{v_eom}
\begin{split}
0 &= \left[ 1 + 2 (b'-\beta) \kappa^2 H^2 \right] \laplace V_k - 2 b \kappa^2 H^2 (Ha)^{-1} \partial_\eta \laplace V_k - \frac{1}{2} \kappa^2 a^{-2} \Xi^{(\text{sr})}_{kl} \partial^l \delta^3(\vec{x}) \\
&\quad+ 2 b \kappa^2 H^2 (Ha)^{-2} \int \left( H(x-x'; \bar{\mu}) + \delta^4(x-x') \ln a \right) \partial^2 \laplace V_k(x') \total^4 x' \eqend{,}
\end{split}
\end{equation}
where we defined the combination
\begin{equation}
\label{kernel_h_def}
H(x-x'; \bar{\mu}) \equiv K(x-y) + K^+(x-y; \bar{\mu}) \eqend{.}
\end{equation}

We see that there are two different contributions to the effective equation of motion: the first one is the classical response of the gravitational field to the spin of the test particle, which comes solely from the Einstein-Hilbert and the point particle action and which consequently is given by all the terms independent of $b$, $b'$ or $\beta$. The second contribution are the quantum corrections due to loops of conformal matter, which are our main interest, and which are sourced by the classical contribution. To see this more explicitly, we split the vector perturbation into a classical and a quantum contribution according to
\begin{equation}
V_k = V_k^\text{cl} + \kappa^2 V_k^\text{qu} \eqend{,}
\end{equation}
and obtain the equations
\begin{equation}
\label{v_eom_cl}
\laplace V_k^\text{cl} = \frac{1}{2} \kappa^2 a^{-2} \Xi^{(\text{sr})}_{kl} \partial^l \delta^3(\vec{x})
\end{equation}
and
\begin{equation}
\label{v_eom_qu}
\begin{split}
\laplace V_k^\text{qu} &= - 2 (b'-\beta) H^2 \laplace V_k^\text{cl} + 2 b H^2 (Ha)^{-1} \partial_\eta \laplace V_k^\text{cl} \\
&\quad- 2 b H^2 (Ha)^{-2} \int \left( H(x-x'; \bar{\mu}) + \delta^4(x-x') \ln a \right) \partial^2 \laplace V_k^\text{cl}(x') \total^4 x' + \bigo{\kappa^2} \eqend{.}
\end{split}
\end{equation}
Since we neglected graviton self-interactions, which would contribute at order $\bigo{\kappa^4}$ in the equation of motion~\eqref{v_eom}, we consequently also have to neglect the $\bigo{\kappa^2}$ correction terms in equation~\eqref{v_eom_qu}.

\subsection{Solutions for the vector perturbation}

Using the well-known formula
\begin{equation}
\laplace \frac{1}{r} = - 4 \pi \delta^3(\vec{x})
\end{equation}
with $r = \abs{\vec{x}}$, the explicit solution of the classical equation~\eqref{v_eom_cl} is given by
\begin{equation}
\label{v_sol_cl}
V_k^\text{cl} = - \frac{1}{8 \pi} \kappa^2 a^{-2} \Xi^{(\text{sr})}_{kl} \partial^l \frac{1}{r} = \frac{\kappa^2}{8 \pi} \Xi^{(\text{sr})}_{kl} \frac{a r^l}{a^3 r^3} \eqend{.}
\end{equation}
Note that it only depends on the physical distance $\hat{r} \equiv a r$ on the equal-time hypersurfaces; therefore, an observer at a fixed physical distance from the source will measure a time-independent Lense-Thirring effect. Since the tensor $\Xi^{(\text{sr})}_{ij}$ is antisymmetric, we can introduce a spin vector $\vec{S}$ according to
\begin{equation}
\label{spin_vector_def}
S^i \equiv \frac{1}{2} \epsilon^{ijk} \Xi^{(\text{sr})}_{jk} \eqend{,}
\end{equation}
such that the solution~\eqref{v_sol_cl} can also be written in the form
\begin{equation}
\label{v_sol_cl_2}
V_k^\text{cl} = \frac{\kappa^2}{8 \pi} \epsilon_{klm} S^m \frac{a r^l}{a^3 r^3} = - \frac{\kappa^2}{8 \pi} \frac{a ( \vec{S} \times \vec{r} )_k}{a^3 r^3} \eqend{.}
\end{equation}

To solve the equation~\eqref{v_eom_qu} for the quantum corrections, we first calculate
\begin{subequations}
\begin{align}
(Ha)^{-1} \partial_\eta \laplace V_k^\text{cl} &= - 2 \laplace V_k^\text{cl} = - \kappa^2 a^{-2} \Xi^{(\text{sr})}_{kl} \partial^l \delta^3(\vec{x}) \eqend{,} \\
\partial^2_\eta \laplace V_k^\text{cl} &= 2 H^2 a^2 \laplace V_k^\text{cl} = \kappa^2 H^2 \Xi^{(\text{sr})}_{kl} \partial^l \delta^3(\vec{x}) \eqend{,}
\end{align}
\end{subequations}
such that equation~\eqref{v_eom_qu}, neglecting the $\bigo{\kappa^2}$ corrections as explained above, reduces to
\begin{equation}
\label{v_eom_qu_2}
\begin{split}
\laplace V_k^\text{qu} &= - 2 (2b+b'-\beta) H^2 \laplace V_k^\text{cl} + 2 b H^2 \ln a \left( 2 - (Ha)^{-2} \laplace \right) \laplace V_k^\text{cl} \\
&\quad- 2 b H^2 (Ha)^{-2} \laplace \int H(x-x'; \bar{\mu}) \laplace V_k^\text{cl}(x') \total^4 x' \\
&\quad+ 4 b H^4 (Ha)^{-2} \int H(x-x'; \bar{\mu}) a^2(x') \laplace V_k^\text{cl}(x') \total^4 x' \\
&= - 2 (2b+b'-\beta) H^2 \laplace V_k^\text{cl} + 2 b H^2 \ln a \left( 2 - (Ha)^{-2} \laplace \right) \laplace V_k^\text{cl} \\
&\quad+ b a^{-2} \kappa^2 H^2 \Xi^{(\text{sr})}_{kl} \partial^l \left[ 2 I_0(x; \bar{\mu}) - (Ha)^{-2} \laplace I_2(x; \bar{\mu}) \right]
\end{split}
\end{equation}
with the integral
\begin{equation}
\label{nonlocal_int}
I_k(x; \bar{\mu}) \equiv a^k(\eta) \int a^{-k}(\eta') \delta^3(\vec{x}') H(x-x'; \bar{\mu}) \total^4 x' \eqend{.}
\end{equation}
The calculation of this integral is somewhat involved and thus relegated to appendix~\ref{sec_appendix}; the result reads~\eqref{appendix_nonlocal_i0},~\eqref{appendix_nonlocal_i2}
\begin{subequations}
\begin{align}
I_0(x; \bar{\mu}) &= - \laplace \frac{\ln \left( \mathe^\gamma \bar{\mu} r \right)}{4 \pi r} \eqend{,} \\
I_2(x; \bar{\mu}) &= - \laplace \frac{\ln \left( \mathe^\gamma \bar{\mu} r \right)}{4 \pi r} - \frac{1}{4 \pi r \eta^2} \eqend{.}
\end{align}
\end{subequations}
Inserting these results into equation~\eqref{v_eom_qu_2} and applying the inverse Laplace operator, we obtain after some rearrangements [using also the explicit solution~\eqref{v_sol_cl} for $V_k^\text{cl}$] the solution
\begin{equation}
\label{v_sol_qu_1}
V_k^\text{qu} = - 2 (3b+b'-\beta) H^2 V_k^\text{cl} - b a^{-2} \kappa^2 H^2 \Xi^{(\text{sr})}_{kl} \left( 2 - (Ha)^{-2} \laplace \right) \partial^l \left( \frac{\ln \left( \mathe^\gamma \bar{\mu} a r \right)}{4 \pi r} \right) \eqend{.}
\end{equation}
This is a well-defined distribution in three spatial dimensions, including the origin $r = 0$. It is, however, not very illuminating, and moreover we know that for very small $r$ the test particle approximation breaks down anyway. We can thus restrict to $r > 0$ and perform the remaining spatial derivatives in the second term using
\begin{equation}
\label{laplace_lnr_r}
\laplace \frac{\ln r}{r} = - \frac{1}{r^3} \qquad (r>0) \eqend{.}
\end{equation}
Neglecting consequently also all local terms proportional to $\delta^3(\vec{x})$ or its derivatives, this gives the result
\begin{equation}
\label{v_sol_qu}
\begin{split}
V_k^\text{qu} &= - (5b+b'-\beta) \frac{\kappa^2 H^2}{4 \pi} \Xi^{(\text{sr})}_{kl} \frac{a r^l}{a^3 r^3} + b \frac{\kappa^2 H^2}{2 \pi} \Xi^{(\text{sr})}_{kl} \frac{a r^l}{a^3 r^3} \ln \left( \mathe^\gamma \bar{\mu} a r \right) + 3 b \kappa^2 \Xi^{(\text{sr})}_{kl} \frac{a r^l}{4 \pi a^5 r^5} \\
&= 2 V_k^\text{cl} \left[ - (5b+b'-\beta) H^2 + 2 b H^2 \ln \left( \mathe^\gamma \bar{\mu} a r \right) + \frac{3 b}{a^2 r^2} \right] \eqend{.}
\end{split}
\end{equation}
Again, the result only depends on the physical distance $\hat{r} = a r$, such that an observer at a fixed physical distance from the particle will see a time-independent quantum correction.

\subsection{Solutions for the scalar perturbations}

Since the action for the scalar potentials $\Phi_\text{A}$ and $\Phi_\text{H}$ is unchanged from the non-spinning case treated previously~\cite{froebverdaguer2016a}, their effective equations of motion and the corresponding solutions are also unchanged. For completeness, we reproduce the solutions here:
\begin{subequations}
\label{phi_result}
\begin{align}
\Phi^\text{cl}_\text{A} &= \Phi^\text{cl}_\text{H} = \frac{\kappa^2 m}{16 \pi a r} \eqend{,} \\
\Phi^\text{qu}_\text{A} &= 2 \Phi^\text{cl}_\text{A} \left[ - H^2 ( 4 b + 3 b' - \beta ) + 2 b H^2 \ln \left( \mathe^\gamma \bar{\mu} a r \right) + \frac{4 b}{3 a^2 r^2} \right] \eqend{,} \\
\Phi^\text{qu}_\text{H} &= 2 \Phi^\text{cl}_\text{H} \left[ - H^2 ( 2 b + 3 b' - \beta ) + 2 b H^2 \ln \left( \mathe^\gamma \bar{\mu} a r \right) + \frac{2 b}{3 a^2 r^2} \right] \eqend{,}
\end{align}
\end{subequations}
and we note that as in the vector case those are valid for $r > 0$.

\section{Comparison with previous results}
\label{sec_comparison}

\subsection{The classical limit}

Plugging the classical contributions~\eqref{v_sol_cl} and~\eqref{phi_result} into the perturbed metric~\eqref{ds_metric_pert} with the decomposition~\eqref{h_munu_inv_gauge} (setting the gauge-dependent vector $X_\mu = 0$), we obtain for the full (linearised) metric the result
\begin{subequations}
\label{result_classical}
\begin{align}
\tilde{g}_{00} &= - a^2 + 2 a^2 \Phi^\text{cl}_\text{A} = - a^2 \left( 1 - \frac{\kappa^2 m}{8 \pi a r} \right) \eqend{,} \\
\tilde{g}_{0k} &= a^2 V^\text{cl}_k = a^2 \frac{\kappa^2}{8 \pi} \Xi^{(\text{sr})}_{kl} \frac{a r^l}{a^3 r^3} \eqend{,} \\
\tilde{g}_{kl} &= a^2 \delta_{kl} + 2 a^2 \delta_{kl} \Phi^\text{cl}_\text{H} = a^2 \delta_{kl} \left( 1 + \frac{\kappa^2 m}{8 \pi a r} \right) \eqend{.}
\end{align}
\end{subequations}
To take the flat-space limit, we first need to transform from conformal time $\eta$ to cosmological time $t$ via
\begin{equation}
\label{eta_t_trafo}
\eta = - H^{-1} \mathe^{- H t} \eqend{,} \qquad \total \eta = \mathe^{- H t} \total t \eqend{,} \qquad a = \mathe^{H t} \eqend{,}
\end{equation}
which results in
\begin{equation}
\begin{split}
\total s^2 &\equiv \tilde{g}_{00} \total \eta^2 + 2 \tilde{g}_{0k} \total \eta \total x^k + \tilde{g}_{kl} \total x^k \total x^l \\
&= a^{-2} \tilde{g}_{00} \total t^2 + 2 a^{-1} \tilde{g}_{0k} \total t \total x^k + \tilde{g}_{kl} \total x^k \total x^l \eqend{.}
\end{split}
\end{equation}
The flat-space limit can now be performed by taking the limit of vanishing Hubble constant $H \to 0$, which entails $a \to 1$ for the scale factor according to equation~\eqref{eta_t_trafo}. Replacing as well $\kappa^2 = 16 \pi G_\text{N}$, we obtain
\begin{equation}
\label{result_flatspace}
\total s^2 = - \left( 1 - \frac{2 G_\text{N} m}{r} \right) \total t^2 + \left( 1 + \frac{2 G_\text{N} m}{r} \right) \total \vec{x}^2 + 4 G_\text{N} \Xi^{(\text{sr})}_{kl} \frac{x^l}{r^3} \total t \total x^k \eqend{.}
\end{equation}
To first order in the mass parameter $M$ and the rotation parameter $\alpha$ (which we employ instead of the more common $a$ to avoid confusion with the de~Sitter scale factor), the Kerr metric~\cite{kerr1963} in Boyer-Lindquist coordinates~\cite{boyerlindquist1967} reads
\begin{equation}
\begin{split}
\total s^2 &= - \left( 1 - \frac{2M}{R} \right) \total t^2 + \left( 1 + \frac{2M}{R} \right) \total R^2 + R^2 \total \theta^2 + R^2 \sin^2 \theta \total \phi^2 - \frac{4 M \alpha \sin^2 \theta}{R} \total t \total \phi \\
&= - \left( 1 - \frac{2M}{r} \right) \total t^2 + \left( 1 + \frac{2M}{r} \right) \total \vec{x}^2 - \frac{4 M \alpha}{r^3} \total t ( x \total y - y \total x ) \eqend{,}
\end{split}
\end{equation}
where $\theta$ and $\phi$ are related in the usual way to the Cartesian coordinates $x$, $y$ and $z$, while we set $R = M + r = M + \sqrt{x^2 + y^2 + z^2}$. Comparing with the flat-space limit~\eqref{result_flatspace}, we identify
\begin{equation}
M = G_\text{N} m \eqend{,} \qquad \Xi^{(\text{sr})}_{xy} = - \Xi^{(\text{sr})}_{yx} = m \alpha \eqend{,}
\end{equation}
with all other components of $\Xi^{(\text{sr})}_{kl}$ vanishing. Note that this is no restriction or loss of generality, since we have just oriented our coordinate system in such a way that the spin vector~\eqref{spin_vector_def} dual to the antisymmetric tensor $\Xi^{(\text{sr})}_{kl}$ points in the $z$ direction. In flat space we thus recover the linearised Kerr solution.

In de~Sitter space, there is a generalised four-dimensional Kerr--de Sitter black hole solution found by Carter~\cite{carter1973} (see also~\cite{gibbonshawking1977,akcaymatzner2011}), and later generalised to $n$ dimensions by Gibbons et\,al.~\cite{gibbonsetal2005}. To first order in the mass parameter $M$ and the rotation parameter $\alpha$, it reads
\begin{equation}
\begin{split}
\total s^2 &= - \left( 1 - H^2 R^2 - \frac{2 M}{R} \right) \total \tau^2 - 2 \alpha \left( H^2 R^2 + \frac{2 M}{R} \right) \sin^2 \theta \total \tau \total \Phi \\
&\quad+ \frac{1}{1 - H^2 R^2} \left( 1 + \frac{2 M}{R (1 - H^2 R^2)} \right) \total R^2 + R^2 \total \theta^2 + R^2 \sin^2 \theta \total \Phi^2 \eqend{.}
\end{split}
\end{equation}
(We note that there seems to be a factor of $\sin^2 \theta$ missing in Ref.~\cite{gibbonshawking1977}.) As $M, \alpha \to 0$, this reduces to the metric of the static patch of de~Sitter spacetime. To obtain the linearised Kerr--de Sitter metric in the Poincar{\'e} patch, one has to perform a coordinate transformation which after some trial and error is obtained in the form
\begin{subequations}
\begin{align}
R &= - \frac{r}{H \eta} + M \eqend{,} \\
\tau &= - \frac{1}{2 H} \ln \left( \eta^2 - r^2 \right) - 2 M \left[ \frac{r \eta}{\eta^2 - r^2} - \frac{1}{2} \ln\left( \frac{r - \eta}{r + \eta} \right) \right] \eqend{,} \\
\Phi &= \phi - \frac{H \alpha}{2} \ln \left( \eta^2 - r^2 \right) + 2 M \alpha H^2 \left[ \frac{\eta}{r} + \frac{r \eta}{r^2 - \eta^2} + \ln\left( \frac{r - \eta}{r + \eta} \right) \right] \eqend{.}
\end{align}
\end{subequations}
This leads to
\begin{equation}
\total s^2 = - a^2 \left( 1 - \frac{2 M}{a r} \right) \total \eta^2 + a^2 \left( 1 + \frac{2 M}{a r} \right) \left( \total r^2 + r^2 \total \theta^2 + r^2 \sin^2 \theta \total \phi^2 \right) - \frac{4 M \alpha}{r} \sin^2 \theta \total \eta \total \phi \eqend{,}
\end{equation}
which is identical to $\tilde{g}_{\mu\nu} \total x^\mu \total x^\nu$ with the metric components~\eqref{result_classical} after a transformation to Cartesian coordinates, with the same identification of parameters as in the flat-space case. We thus recover a linearised Kerr solution also in de~Sitter space.

\subsection{The flat-space limit}

The quantum corrections for spinning particles in flat space, due to loops of gauge fields and massless and massive fermions and scalars with arbitrary curvature coupling, have been recently studied by one of us~\cite{froeb2016}. The result in the massless case is given by
\begin{subequations}
\begin{align}
\Phi_\text{A} &= \frac{\kappa^2 M}{16 \pi r} \left[ 1 + \left[ N_0 \left( 1 + \frac{5}{4} (1-6\xi)^2 \right) + 6 N_{1/2} + 12 N_1 \right] \frac{\kappa^2}{720 \pi^2 r^2} \right] \eqend{,} \\
\Phi_\text{H} &= \frac{\kappa^2 M}{16 \pi r} \left[ 1 + \left[ N_0 \left( 1 - \frac{5}{2} (1-6\xi)^2 \right) + 6 N_{1/2} + 12 N_1 \right] \frac{\kappa^2}{1440 \pi^2 r^2} \right] \eqend{,} \\
V_i &= - \frac{\kappa^2 (\vec{S} \times \vec{r})_i}{8 \pi r^3} \left[ 1 + \left( N_0 + 6 N_{1/2} + 12 N_1 \right) \frac{\kappa^2}{320 \pi^2 r^2} \right] \eqend{,}
\end{align}
\end{subequations}
where $N_s$ denotes the number of free, massless fields of spin $s$, and where the (constant) spin vector $\vec{S}$ is given by equation~\eqref{spin_vector_def}. Conformal coupling for the scalars entails $\xi = 1/6$, and comparing with the expressions~\eqref{coeffs_b} that determine the parameters $b$ and $b'$ in the free-field case, the result of Ref.~\cite{froeb2016} can be written as
\begin{subequations}
\label{result_gaugeinv}
\begin{align}
\Phi_\text{A} &= \frac{\kappa^2 M}{16 \pi r} \left[ 1 + \kappa^2 \frac{8 b}{3 r^2} \right] \eqend{,} \\
\Phi_\text{H} &= \frac{\kappa^2 M}{16 \pi r} \left[ 1 + \kappa^2 \frac{4 b}{3 r^2} \right] \eqend{,} \\
V_i &= \frac{\kappa^2 \Xi^{(\text{sr})}_{ij} x^j}{8 \pi r^3} \left[ 1 + \kappa^2 \frac{6 b}{r^2} \right] \eqend{,}
\end{align}
\end{subequations}
which coincides exactly with our results~\eqref{v_sol_cl}, \eqref{v_sol_qu} and~\eqref{phi_result} in the flat-space limit $a \to 1$, $H \to 0$.

We can also compare with the (corrected) results of Bjerrum-Bohr, Donoghue and Holstein~\cite{bjerrumbohrdonoghueholstein2003a} and Khriplovich and Kirillin~\cite{khriplovichkirilin2003}, who studied quantum corrections due to loops of gravitons. Both use harmonic gauge, better known as de~Donder gauge, which is determined by the condition
\begin{equation}
\partial_\mu \left( h^{\mu\nu} - \frac{1}{2} \eta^{\mu\nu} h \right) = 0 \eqend{.}
\end{equation}
In terms of the decomposition of the metric perturbations into gauge-invariant and gauge-dependent parts~\eqref{h_munu_inv_gauge} (taking the flat-space limit), this gives rise to
\begin{subequations}
\begin{align}
\partial^2 X_0 &= \partial_\eta \left( \Phi_\text{A} + 3 \Phi_\text{H} \right) \eqend{,} \\
\partial^2 X_j &= \partial_\eta V_j - \partial_j \Phi_\text{A} + \partial_j \Phi_\text{H} \eqend{,}
\end{align}
\end{subequations}
which determines the vector $X^\mu$ in terms of the gauge-invariant potentials. Since the flat-space result~\eqref{result_gaugeinv} is time-independent, we also assume a time-independent vector $X_\mu$ in order to avoid introducing a spurious time dependence in our results. We then obtain $X_0 = 0$ and, using equation~\eqref{laplace_lnr_r},
\begin{equation}
X_j = \frac{\partial_j}{\laplace} \left( \Phi_\text{H} - \Phi_\text{A} \right) = \frac{\kappa^4 M b}{12 \pi} \partial_j \left( \frac{\ln r}{r} \right) = \frac{\kappa^4 M b}{12 \pi} \frac{x_j (1-\ln r)}{r^3} \eqend{.}
\end{equation}
In this gauge, our result for the metric perturbation thus reads
\begin{subequations}
\begin{align}
h_{00} &= 2 \Phi_\text{A} = \frac{\kappa^2 M}{8 \pi r} \left[ 1 + \kappa^2 \frac{8 b}{3 r^2} \right] \eqend{,} \\
h_{0i} &= V_i + 2 \partial_i X_0 = \frac{\kappa^2 \Xi^{(\text{sr})}_{ij} x^j}{8 \pi r^3} \left[ 1 + \kappa^2 \frac{6 b}{r^2} \right] \eqend{,} \\
h_{ij} &= 2 \delta_{ij} \Phi_\text{H} + 2 \partial_{(i} X_{j)} = \delta_{ij} \frac{\kappa^2 M}{8 \pi r} \left[ 1 + \kappa^2 \frac{4 b}{3 r^2} (2-\ln r) \right] + \frac{\kappa^4 M b}{6 \pi} x_i x_j \frac{(-4+3\ln r)}{r^5} \eqend{,}
\end{align}
\end{subequations}
which has the structure of the results of Refs.~\cite{bjerrumbohrdonoghueholstein2003a,khriplovichkirilin2003} (including the logarithmic terms), but with different numerical coefficients since their results include corrections due to graviton loops.

\section{Discussion}
\label{sec_discussion}

Quantum effects in de~Sitter spacetime play an important role in cosmology, not only because de~Sitter approximates very well most of the inflationary period in the standard cosmological model~\cite{mukhanov,kazanas1980,sato1981,guth1981,linde1982,albrechtsteinhardt1982}, but also because this spacetime models our present accelerated universe~\cite{perlmutteretal1998,knopetal2003,eisensteinetal2005,astieretal2006,kowalskietal2008}. For this reason, it is important to test perturbative quantum field theory in a de~Sitter background (see, e.g., the recent Refs.~\cite{gazeaulachieze2006,higuchi2008,brosepsteinmoschella2010,higuchimarolfmorrison2011a,marolfmorrison2011a,marolfmorrison2011b,hollands2013,hollands2012,marolfmorrisonsrednicki2013,koraitanaka2013,moschella2016,belokognefolacciqueva2016} and references therein) as well as perturbative quantum field theory interacting with metric perturbations, both quantized and classical, at tree level and beyond (see, e.g., the recent Refs.~\cite{miaowoodard2006,kahyawoodard2007,higuchimarolfmorrison2011b,moratsamiswoodard2012,frv2012,fprv2013,frv2014,battistaesposito2014,battistaetal2015,parkprokopecwoodard2016,wangwoodard2015a,battistaetal2017} and references therein), even if the effects are too small to be observed at present~\cite{weinberg2005}. The present work is another contribution to this large research field, probing the vector part of the metric perturbations which is often neglected. In the usual scalar-driven inflationary models, vector perturbations are not generated on large scales, and any existing perturbations are quickly redshifted and negligible at late times~\cite{mukhanov}. In our case, even though inflation is driven by a cosmological constant, the spinning point particle continously excites vector modes of the metric perturbation, which consequently remain present even at late times.

We have calculated the quantum correction to the gauge-invariant vector mode of the metric perturbation in de~Sitter space, due to the vacuum fluctuations of conformal matter. As in our previous work~\cite{froebverdaguer2016a} where we calculated corrections to the Bardeen potentials, our result is valid for arbitrary conformal field theories (including strongly interacting ones). It depends on the parameters $b$ and $b'$ which appear in the trace anomaly, and which are given by equation~\eqref{coeffs_b} for free theories, but generally take different values once interactions are included. Reintroducing $\hbar$ and $c$, and using the Planck length $\ell_\text{Pl} = \sqrt{\hbar G_\text{N}/c^3}$ and the physical distance on equal-time hypersurfaces $\hat{r} \equiv a r$, we can write it in the form
\begin{equation}
\vec{V} = - 2 G_\text{N} \frac{\vec{S} \times \hat{\vec{r}}}{\hat{r}^3} \bigg[ 1 + 96 \pi b \frac{\ell_\text{Pl}^2}{\hat{r}^2} + 32 \pi \ell_\text{Pl}^2 H^2 \left( \beta - 5b - b' + 2 c(\mu) + 2 b \ln \left( \mathe^\gamma \mu \hat{r} \right) \right) \bigg] \eqend{.}
\end{equation}
Both in the flat-space limit $H \to 0$ and the classical limit $\ell_\text{Pl} \to 0$, this reproduces previous results for the quantum corrections and for the (linearised) Kerr(-de~Sitter) metric. We see that there are three different contributions to the quantum corrections: a) one which is also present in flat space, independent of the Hubble constant $H$, and which only is significant at distances of the order of the Planck length, b) a constant correction, which depends on the renormalized parameters appearing in front of the term quadratic in the curvature tensors of the gravitational action and the renormalisation scale, and which could be absorbed in a renormalisation of the spin of the point particle, and c) a correction that grows logarithmically with the distance from the particle. This last term is the most interesting, since while the prefactor $\ell_\text{Pl}^2 H^2$ is extremely small at present times, during inflation it is small but appreciable. It is thus conceivable that the logarithmic growth at large distances $\hat{r}$ could overcome the smallness of this prefactor and have potentially observable effects. However, this growth might be an artifact of perturbation theory, similar to the infrared growth of loop corrections for the massless, minimally coupled scalar field with $\phi^4$ interaction. These loop corrections can be resummed~\cite{starobinsky1994,woodard2005,riottosloth2008,rajaraman2010,benekemoch2013,serreauparentani2013,gautierserreau2013,youssefkreimer2014,lopeznaciretal2016} to obtain a non-perturbative result without infrared growth, but with strongly non-Gaussian behaviour. We thus perform a tentative ``resummation'' of the logarithmic term according to
\begin{equation}
\frac{1}{\hat{r}^3} \bigg[ 1 + 64 \pi \ell_\text{Pl}^2 H^2 b \ln \left( \mathe^\gamma \mu \hat{r} \right) \bigg] = \frac{1}{\hat{r}^{3 - 64 \pi b \ell_\text{Pl}^2 H^2}} + \bigo{\ell_\text{Pl}^4}
\end{equation}
into a modified power-law to order $\ell_\text{Pl}^2$, to which we are working. Thus, since $b > 0$, which can be seen from the explicit expression~\eqref{coeffs_b} for free theories, and is also true for interacting theories because of unitarity~\cite{osbornpetkou1994}, the potential decays slower at large distances. The gravitational force is thus enhanced with respect to the classical case, in the same way as for the Bardeen potentials~\cite{froebverdaguer2016a}.\footnote{See footnote~\ref{errnote} on page~\pageref{errnote}.}

The above results have been obtained for conformal matter fields, where the quantum correction only depends on the physical distance $\hat{r} = a r$ and is thus time-independent for observers at a fixed physical distance from the particle. For quantum corrections due to other matter fields, this need not be the case, and existing calculations~\cite{parkprokopecwoodard2016} indicate that one might expect contributions which grow at late times like $\ln a$. In particular, such contributions might be expected if one considers the vacuum fluctuations of a massless, minimally coupled scalar field, and it would be important to generalise our calculation to this case.

\begin{acknowledgments}
E.~V.\ acknowledges partial financial support from the Research Projects FPA2013-46570-C2-2-P, AGAUR 2014-SGR-1474, MDM-2014-0369 of ICCUB (Unidad de Excelencia `Mar{\'\i}a de Maeztu') and CPAN CSD2007-00042, with\-in the program Consolider-Ingenio 2010. This work is part of a project that has received funding from the European Union’s Horizon 2020 research and innovation programme under the Marie Sk{\l}odowska-Curie grant agreement No.~702750 ``QLO-QG''.
\end{acknowledgments}

\appendix

\section{Calculation of the non-local term}
\label{sec_appendix}

In this appendix, we calculate the integral from equation~\eqref{nonlocal_int}. Performing a Fourier transform, we get
\begin{equation}
\begin{split}
I_k(x; \bar{\mu}) &= a^k(\eta) \int a^{-k}(\eta') \delta^3(\vec{x}') H(x-x'; \bar{\mu}) \total^4 x' \\
&\quad= \int \left[ \int \left( \frac{\eta'}{\eta} \right)^k \tilde{H}(\eta-\eta', \vec{p}; \bar{\mu}) \total \eta' \right] \mathe^{\mathi \vec{p} \vec{x}} \frac{\total^3 p}{(2\pi)^3} \equiv \int \tilde{I}_k(\eta, \vec{p}; \bar{\mu}) \mathe^{\mathi \vec{p} \vec{x}} \frac{\total^3 p}{(2\pi)^3} \eqend{.}
\end{split}
\end{equation}
Inserting the Fourier transforms of the kernels $K$~\eqref{kernels_def} into the definition of the kernel $H(x-x'; \bar{\mu})$~\eqref{kernel_h_def} and performing the integral over $p^0$, we obtain that~\cite{fprv2013,froebverdaguer2016a}
\begin{equation}
\label{kernel_h_fourier}
\tilde{H}(\eta-\eta', \vec{p}; \bar{\mu}) = \distlim_{\epsilon \to 0} \cos\left[ \abs{\vec{p}} (\eta-\eta') \right] \left[ \frac{\Theta(\eta-\eta'-\epsilon)}{\eta-\eta'} + \delta(\eta-\eta') \left( \ln(\bar{\mu} \epsilon) + \gamma \right) \right] \eqend{,}
\end{equation}
where the notation $\distlim$ means that the limit has to be taken in the sense of distributions, i.e., after integrating. It is clearly seen that the integral over $\eta'$ of this kernel is not convergent if $k \geq 0$. The physical reason for this is that we evolve the quantum system starting from a free vacuum state (the Bunch-Davies vacuum), with the implicit assumption (as in flat space) that in the far past, the particles become free and the effective interaction between them vanishes. However, as $\eta' \to -\infty$, the universe shrinks and interactions between particles can only become stronger, such that we cannot assume a free state in the past. There are two solutions to this problem: one could either start at a finite initial time $\eta' = \eta_0$ and include perturbative corrections to the initial state at this time~\cite{collinsholman2005,collins2013,collinsholmanverdanyan2014}, or employ an $\mathi \epsilon$ prescription to select an adiabatic interacting vacuum state at past infinity~\cite{peskinschroeder,marolfmorrison2011a,higuchimarolfmorrison2011a,marolfmorrison2011b,frv2012}. Both solutions are expected to agree at least in the late-time limit $\eta \to 0$, and for ease of implementation we employ the second one. As explained in Ref.~\cite{froebverdaguer2016a}, the net effect of the $\mathi \epsilon$ prescription is to multiply the spatial Fourier transform~\eqref{kernel_h_fourier} by a factor $\exp\left[ - \mathi \epsilon \abs{\vec{p}} (\eta-\eta') \right]$. In order not to confuse the two parameters $\epsilon$ (one coming from the proper definition of the distribution~\eqref{kernel_h_fourier} and one selecting the adiabatic interacting vacuum state), we denote the prescription parameter by $\delta$, and obtain
\begin{equation}
\begin{split}
\tilde{I}_k(\eta, \vec{p}; \bar{\mu}) = \lim_{\epsilon \to 0} \mathe^{ - \delta \abs{\vec{p}} \eta } \int &\mathe^{ \delta \abs{\vec{p}} \eta' } \left( \frac{\eta'}{\eta} \right)^k \cos\left[\abs{\vec{p}} (\eta-\eta')\right] \\
&\quad\times \left[ \frac{\Theta(\eta-\eta'-\epsilon)}{\eta-\eta'} + \delta(\eta-\eta') \left( \ln(\bar{\mu} \epsilon) + \gamma \right) \right] \total \eta' \eqend{,}
\end{split}
\end{equation}
with $\delta > 0$. The second part including the $\delta$ distribution is of course easily solved; for the other one we introduce an initial time $\eta_0$, express the cosine with exponentials and obtain
\begin{equation}
\tilde{I}_k(\eta, \vec{p}; \bar{\mu}) = \mathe^{ - \delta \abs{\vec{p}} \eta } \lim_{\epsilon \to 0} \left[ \Re \int_{\eta_0}^{\eta-\epsilon} \left( \frac{\eta'}{\eta} \right)^k \frac{\mathe^{\mathi \abs{\vec{p}} (\eta-\eta') + \delta \abs{\vec{p}} \eta'}}{\eta-\eta'} \total \eta' + \ln(\bar{\mu} \epsilon) + \gamma \right] \eqend{.}
\end{equation}
We then decompose
\begin{equation}
\left( \frac{\eta'}{\eta} \right)^k \frac{1}{\eta-\eta'} = \frac{1}{\eta-\eta'} - \frac{1}{\eta} \sum_{m=0}^{k-1} \left( \frac{\eta'}{\eta} \right)^m
\end{equation}
such that
\begin{equation}
\begin{split}
\tilde{I}_k(\eta, \vec{p}; \bar{\mu}) = \mathe^{ - \delta \abs{\vec{p}} \eta } \lim_{\epsilon \to 0} \Re \Bigg[ &\int_{\eta_0}^{\eta-\epsilon} \frac{\mathe^{\mathi \abs{\vec{p}} (\eta-\eta') + \delta \abs{\vec{p}} \eta'}}{\eta-\eta'} \total \eta' \\
&\quad- \frac{1}{\eta} \sum_{m=0}^{k-1} (\abs{\vec{p}} \eta)^{-m} \frac{\partial^m}{\partial \delta^m} \int_{\eta_0}^{\eta-\epsilon} \mathe^{\mathi \abs{\vec{p}} (\eta-\eta') + \delta \abs{\vec{p}} \eta'} \total \eta' + \ln(\bar{\mu} \epsilon) + \gamma \Bigg] \eqend{.}
\end{split}
\end{equation}
Using the indefinite integral
\begin{equation}
\int \frac{\mathe^{a x}}{x-x_0} \total x = \mathe^{a x_0} \left[ \Ein[a(x-x_0)] + \ln(x-x_0) \right]
\end{equation}
we thus obtain
\begin{equation}
\begin{split}
&\tilde{I}_k(\eta, \vec{p}; \bar{\mu}) = \lim_{\epsilon \to 0} \Re \Bigg[ \Ein[ (\mathi-\delta) \abs{\vec{p}} (\eta-\eta_0)] + \ln(\eta-\eta_0) - \Ein[ (\mathi-\delta) \abs{\vec{p}} \epsilon] - \ln(\epsilon) \\
&\qquad+ \sum_{m=0}^{k-1} (\abs{\vec{p}} \eta)^{-m-1} \sum_{n=0}^m \frac{m!}{(m-n)!} \frac{\abs{\vec{p}}^{m-n}}{(\mathi-\delta)^{n+1}} \left[ \mathe^{(\mathi-\delta) \abs{\vec{p}} \epsilon} (\eta-\epsilon)^{m-n} - \mathe^{(\mathi-\delta) \abs{\vec{p}} (\eta-\eta_0)} \eta_0^{m-n} \right] \\
&\qquad+ \mathe^{ - \delta \abs{\vec{p}} \eta } \left[ \ln(\bar{\mu} \epsilon) + \gamma \right] \Bigg] \eqend{.}
\end{split}
\end{equation}
The terms depending on $\eta_0$ must be absorbed in a correction to the initial state if $\delta = 0$. In our case, we want to select an interacting vacuum state at past infinity using the $\mathi \delta$ prescription, and thus take first the initial time to past infinity, $\eta_0 \to -\infty$ and afterwards the limit $\delta \to 0$. Finally, we can then also take the limit $\epsilon \to 0$ coming from the proper definition of the kernel $H(x-x'; \bar{\mu})$ as a distribution~\eqref{kernel_h_fourier}. For the asymptotic expansion of the $\Ein$ special function we have~\cite{frv2012}
\begin{equation}
\begin{split}
\Ein\left[ (\mathi-\delta) \abs{\vec{p}} (\eta-\eta_0) \right] &\sim - \gamma - \ln\left[ - (\mathi-\delta) \abs{\vec{p}} (\eta-\eta_0) \right] \\
&\qquad+ \mathe^{(\mathi-\delta) \abs{\vec{p}} (\eta-\eta_0)} \left[ \frac{1}{- (\mathi-\delta) \abs{\vec{p}} \eta_0} + \bigo{\eta_0^{-2}} \right] \eqend{,}
\end{split}
\end{equation}
and since $\Ein(0) = 0$ we get
\begin{equation}
\tilde{I}_k(\eta, \vec{p}; \bar{\mu}) = \ln \frac{\bar{\mu}}{\abs{\vec{p}}} + \Re \sum_{m=0}^{k-1} \sum_{n=0}^m \frac{m!}{(m-n)!} (\abs{\vec{p}} \eta)^{-n-1} (-\mathi)^{n+1} \eqend{.}
\end{equation}
For $k = 1$, this integral is the same as in Ref.~\cite{froebverdaguer2016a}, but we need it for $k = 0$ and $k = 2$ where we get
\begin{equation}
\tilde{I}_0(\eta, \vec{p}; \bar{\mu}) = \ln \frac{\bar{\mu}}{\abs{\vec{p}}} \eqend{,} \quad \tilde{I}_2(\eta, \vec{p}; \bar{\mu}) = \ln \frac{\bar{\mu}}{\abs{\vec{p}}} - (\abs{\vec{p}} \eta)^{-2} \eqend{.}
\end{equation}
The reverse Fourier transform of the first integral was already done in that reference as well, and reads
\begin{equation}
\label{appendix_nonlocal_i0}
I_0(x; \bar{\mu}) = - \laplace \frac{\ln \left( \mathe^\gamma \bar{\mu} r \right)}{4 \pi r} \eqend{,}
\end{equation}
which is a well-defined distribution. For the second part, we calculate
\begin{equation}
\int \abs{\vec{p}}^{-2} \mathe^{\mathi \vec{p} \vec{x}} \frac{\total^3 p}{(2\pi)^3} = \frac{1}{2 \pi^2} \lim_{\epsilon \to 0} \int_0^\infty \mathe^{-\epsilon p} \frac{\sin(p r)}{p r} \total p = \frac{1}{2 \pi^2} \lim_{\epsilon \to 0} \int_0^\infty \mathe^{-\epsilon p} \sum_{k=0}^\infty \frac{(-p^2 r^2)^k}{(2k+1)!} \total p
\end{equation}
and use that for $\Re a < 0$ and $b \geq 0$ we have
\begin{equation}
\int_0^\infty \mathe^{a p} p^b \total p = \frac{\Gamma(b+1)}{(-a)^{b+1}}
\end{equation}
to obtain
\begin{equation}
\int \abs{\vec{p}}^{-2} \mathe^{\mathi \vec{p} \vec{x}} \frac{\total^3 p}{(2\pi)^3} = \frac{1}{2 \pi^2} \lim_{\epsilon \to 0} \sum_{k=0}^\infty \frac{(-r^2)^k}{(2k+1) \epsilon^{2k+1}} = \frac{1}{2 \pi^2 r} \lim_{\epsilon \to 0} \arctan \frac{r}{\epsilon} = \frac{1}{4 \pi r} \eqend{,}
\end{equation}
such that
\begin{equation}
\label{appendix_nonlocal_i2}
I_2(x; \bar{\mu}) = - \laplace \frac{\ln \left( \mathe^\gamma \bar{\mu} r \right)}{4 \pi r} - \frac{1}{4 \pi r \eta^2} \eqend{.}
\end{equation}

\bibliography{literature}

\providecommand{\href}[2]{#2}\begingroup\raggedright\begin{thebibliography}{100}

\bibitem{donoghue1994b}
J.~F. Donoghue, \emph{{General relativity as an effective field theory: The
  leading quantum corrections}},
  \href{http://dx.doi.org/10.1103/PhysRevD.50.3874}{\emph{Phys.~Rev.~D} {\bf
  50} (1994) 3874}, [\href{http://arxiv.org/abs/gr-qc/9405057}{{\tt
  gr-qc/9405057}}].

\bibitem{burgess2004}
C.~P. Burgess, \emph{{Quantum gravity in everyday life: General relativity as
  an effective field theory}},
  \href{http://dx.doi.org/10.12942/lrr-2004-5}{\emph{Living~Rev.~Rel.} {\bf 7}
  (2004) 5}, [\href{http://arxiv.org/abs/gr-qc/0311082}{{\tt gr-qc/0311082}}].

\bibitem{radkowski1970}
A.~F. Radkowski, \emph{{Some Aspects of the Source Description of
  Gravitation}},
  \href{http://dx.doi.org/10.1016/0003-4916(70)90021-7}{\emph{Ann.~Phys.} {\bf
  56} (1970) 319}.

\bibitem{schwinger1968}
J.~S. Schwinger, \emph{{Sources and gravitons}},
  \href{http://dx.doi.org/10.1103/PhysRev.173.1264}{\emph{Phys.~Rev.} {\bf 173}
  (1968) 1264}.

\bibitem{duff1974}
M.~J. Duff, \emph{{Quantum corrections to the Schwarzschild solution}},
  \href{http://dx.doi.org/10.1103/PhysRevD.9.1837}{\emph{Phys.~Rev.~D} {\bf 9}
  (1974) 1837}.

\bibitem{capperduffhalpern1974}
D.~M. Capper, M.~J. Duff and L.~Halpern, \emph{{Photon corrections to the
  graviton propagator}},
  \href{http://dx.doi.org/10.1103/PhysRevD.10.461}{\emph{Phys.~Rev.~D} {\bf 10}
  (1974) 461}.

\bibitem{capperduff1974}
D.~M. Capper and M.~J. Duff, \emph{{The one-loop neutrino contribution to the
  graviton propagator}},
  \href{http://dx.doi.org/10.1016/0550-3213(74)90582-3}{\emph{Nucl.~Phys.~B}
  {\bf 82} (1974) 147}.

\bibitem{donoghue1994a}
J.~F. Donoghue, \emph{{Leading quantum correction to the Newtonian potential}},
  \href{http://dx.doi.org/10.1103/PhysRevLett.72.2996}{\emph{Phys.~Rev.~Lett.}
  {\bf 72} (1994) 2996}, [\href{http://arxiv.org/abs/gr-qc/9310024}{{\tt
  gr-qc/9310024}}].

\bibitem{muzinichvokos1995}
I.~J. Muzinich and S.~Vokos, \emph{{Long range forces in quantum gravity}},
  \href{http://dx.doi.org/10.1103/PhysRevD.52.3472}{\emph{Phys.~Rev.~D} {\bf
  52} (1995) 3472}, [\href{http://arxiv.org/abs/hep-th/9501083}{{\tt
  hep-th/9501083}}].

\bibitem{hamberliu1995}
H.~W. Hamber and S.~Liu, \emph{{On the quantum corrections to the Newtonian
  potential}},
  \href{http://dx.doi.org/10.1016/0370-2693(95)00790-R}{\emph{Phys.~Lett.~B}
  {\bf 357} (1995) 51}, [\href{http://arxiv.org/abs/hep-th/9505182}{{\tt
  hep-th/9505182}}].

\bibitem{akhundovbelluccishiekh1997}
A.~A. Akhundov, S.~Bellucci and A.~Shiekh, \emph{{Gravitational interaction to
  one loop in effective quantum gravity}},
  \href{http://dx.doi.org/10.1016/S0370-2693(96)01694-2}{\emph{Phys.~Lett.~B}
  {\bf 395} (1997) 16}, [\href{http://arxiv.org/abs/gr-qc/9611018}{{\tt
  gr-qc/9611018}}].

\bibitem{dalvitmazzitelli1997}
D.~A.~R. Dalvit and F.~D. Mazzitelli, \emph{{Geodesics, gravitons and the gauge
  fixing problem}},
  \href{http://dx.doi.org/10.1103/PhysRevD.56.7779}{\emph{Phys.~Rev.~D} {\bf
  56} (1997) 7779}, [\href{http://arxiv.org/abs/hep-th/9708102}{{\tt
  hep-th/9708102}}].

\bibitem{duffliu2000a}
M.~J. Duff and J.~T. Liu, \emph{{Complementarity of the Maldacena and
  Randall-Sundrum pictures}},
  \href{http://dx.doi.org/10.1088/0264-9381/18/16/310}{\emph{Class.~Quant.~Grav.}
  {\bf 18} (2001) 3207}, [\href{http://arxiv.org/abs/hep-th/0003237}{{\tt
  hep-th/0003237}}].

\bibitem{duffliu2000b}
M.~J. Duff and J.~T. Liu, \emph{{Complementarity of the Maldacena and
  Randall-Sundrum pictures}},
  \href{http://dx.doi.org/10.1103/PhysRevLett.85.2052}{\emph{Phys.~Rev.~Lett.}
  {\bf 85} (2000) 2025}, [\href{http://arxiv.org/abs/hep-th/0003237}{{\tt
  hep-th/0003237}}].

\bibitem{kirilinkhriplovich2002}
G.~G. Kirilin and I.~B. Khriplovich, \emph{{Quantum power correction to the
  Newton law}},
  \href{http://dx.doi.org/10.1134/1.1537290}{\emph{J.~Exp.~Theor.~Phys.} {\bf
  95} (2002) 981}, [\href{http://arxiv.org/abs/gr-qc/0207118}{{\tt
  gr-qc/0207118}}]. {[Zh.~Eksp.~Teor.~Fiz. \textbf{95} (2002) 1139]}.

\bibitem{khriplovichkirilin2003}
I.~B. Khriplovich and G.~G. Kirilin, \emph{{Quantum long range interactions in
  general relativity}},
  \href{http://dx.doi.org/10.1134/1.1777618}{\emph{J.~Exp.~Theor.~Phys.} {\bf
  98} (2004) 1063}, [\href{http://arxiv.org/abs/gr-qc/0402018}{{\tt
  gr-qc/0402018}}]. {[Zh.~Eksp.~Teor.~Fiz. \textbf{125} (2004) 1219]}.

\bibitem{bjerrumbohrdonoghueholstein2003a}
N.~E.~J. Bjerrum-Bohr, J.~F. Donoghue and B.~R. Holstein, \emph{{Quantum
  corrections to the Schwarzschild and Kerr metrics}},
  \href{http://dx.doi.org/10.1103/PhysRevD.68.084005}{\emph{Phys.~Rev.~D} {\bf
  68} (2003) 084005}, [\href{http://arxiv.org/abs/hep-th/0211071}{{\tt
  hep-th/0211071}}]. Erratum:
  \href{http://dx.doi.org/10.1103/PhysRevD.71.069904}{\emph{Phys.~Rev.~D}
  \textbf{71} (2005) 069904}.

\bibitem{bjerrumbohrdonoghueholstein2003b}
N.~E.~J. Bjerrum-Bohr, J.~F. Donoghue and B.~R. Holstein, \emph{{Quantum
  gravitational corrections to the nonrelativistic scattering potential of two
  masses}},
  \href{http://dx.doi.org/10.1103/PhysRevD.67.084033}{\emph{Phys.~Rev.~D} {\bf
  67} (2003) 084033}, [\href{http://arxiv.org/abs/hep-th/0211072}{{\tt
  hep-th/0211072}}]. Erratum:
  \href{http://dx.doi.org/10.1103/PhysRevD.71.069903}{\emph{Phys.~Rev.~D}
  \textbf{71} (2005) 069903}.

\bibitem{satzmazzitellialvarez2005}
A.~Satz, F.~D. Mazzitelli and E.~Alvarez, \emph{{Vacuum polarization around
  stars: Nonlocal approximation}},
  \href{http://dx.doi.org/10.1103/PhysRevD.71.064001}{\emph{Phys.~Rev.~D} {\bf
  71} (2005) 064001}, [\href{http://arxiv.org/abs/gr-qc/0411046}{{\tt
  gr-qc/0411046}}].

\bibitem{holsteinross2008}
B.~R. Holstein and A.~Ross, \emph{{Spin Effects in Long Range Gravitational
  Scattering}},  \href{http://arxiv.org/abs/0802.0716}{{\tt 0802.0716}}.

\bibitem{parkwoodard2010}
S.~Park and R.~P. Woodard, \emph{{Solving the Effective Field Equations for the
  Newtonian Potential}},
  \href{http://dx.doi.org/10.1088/0264-9381/27/24/245008}{\emph{Class.~Quant.~Grav.}
  {\bf 27} (2010) 245008}, [\href{http://arxiv.org/abs/1007.2662}{{\tt
  1007.2662}}].

\bibitem{marunovicprokopec2011}
A.~Marunovic and T.~Prokopec, \emph{{Time transients in the quantum corrected
  Newtonian potential induced by a massless nonminimally coupled scalar
  field}},
  \href{http://dx.doi.org/10.1103/PhysRevD.83.104039}{\emph{Phys.~Rev.~D} {\bf
  83} (2011) 104039}, [\href{http://arxiv.org/abs/1101.5059}{{\tt 1101.5059}}].

\bibitem{marunovicprokopec2012}
A.~Marunovic and T.~Prokopec, \emph{{Antiscreening in perturbative quantum
  gravity and resolving the Newtonian singularity}},
  \href{http://dx.doi.org/10.1103/PhysRevD.87.104027}{\emph{Phys.~Rev.~D} {\bf
  87} (2013) 104027}, [\href{http://arxiv.org/abs/1209.4779}{{\tt 1209.4779}}].

\bibitem{burnspilaftsis2015}
D.~Burns and A.~Pilaftsis, \emph{{Matter Quantum Corrections to the Graviton
  Self-Energy and the Newtonian Potential}},
  \href{http://dx.doi.org/10.1103/PhysRevD.91.064047}{\emph{Phys.~Rev.~D} {\bf
  91} (2015) 064047}, [\href{http://arxiv.org/abs/1412.6021}{{\tt 1412.6021}}].

\bibitem{bjerrumbohretal2016}
N.~E.~J. Bjerrum-Bohr, J.~F. Donoghue, B.~R. Holstein, L.~Plante and
  P.~Vanhove, \emph{{Light-like Scattering in Quantum Gravity}},
  \href{http://dx.doi.org/10.1007/JHEP11(2016)117}{\emph{JHEP} {\bf 11} (2016)
  117}, [\href{http://arxiv.org/abs/1609.07477}{{\tt 1609.07477}}].

\bibitem{mukhanov}
V.~Mukhanov, \emph{{Physical Foundations of Cosmology}}.
\newblock \href{http://www.worldcat.org/search?q=isbn:9780521563987}{Cambridge University Press, Cambridge, UK, 2005}.

\bibitem{kazanas1980}
D.~Kazanas, \emph{{Dynamics of the Universe and Spontaneous Symmetry
  Breaking}}, \href{http://dx.doi.org/10.1086/183361}{\emph{Astrophys.~J.} {\bf
  241} (1980) L59}.

\bibitem{sato1981}
K.~Sato, \emph{{First order phase transition of a vacuum and expansion of the
  Universe}},
  \href{http://dx.doi.org/https://doi.org/10.1093/mnras/195.3.467}{\emph{Mon.~Not.~Roy.~Astron.~Soc.}
  {\bf 195} (1981) 467}.

\bibitem{guth1981}
A.~H. Guth, \emph{{The Inflationary Universe: a possible solution to the
  horizon and flatness problems}},
  \href{http://dx.doi.org/10.1103/PhysRevD.23.347}{\emph{Phys.~Rev.~D} {\bf 23}
  (1981) 347}.

\bibitem{linde1982}
A.~D. Linde, \emph{{A new inflationary universe scenario: a possible solution
  of the horizon, flatness, homogeneity, isotropy, and primordial monopole
  problems}},
  \href{http://dx.doi.org/10.1016/0370-2693(82)91219-9}{\emph{Phys.~Lett.~B}
  {\bf 108} (1982) 389}.

\bibitem{albrechtsteinhardt1982}
A.~Albrecht and P.~J. Steinhardt, \emph{{Cosmology for grand unified theories
  with radiatively induced symmetry breaking}},
  \href{http://dx.doi.org/10.1103/PhysRevLett.48.1220}{\emph{Phys.~Rev.~Lett.}
  {\bf 48} (1982) 1220}.

\bibitem{perlmutteretal1998}
{\scshape Supernova Cosmology Project} collaboration, S.~Perlmutter et~al.,
  \emph{{Measurements of $\Omega$ and $\Lambda$ from 42 high redshift
  supernovae}}, \href{http://dx.doi.org/10.1086/307221}{\emph{Astrophys.~J.}
  {\bf 517} (1999) 565}, [\href{http://arxiv.org/abs/astro-ph/9812133}{{\tt
  astro-ph/9812133}}].

\bibitem{knopetal2003}
{\scshape Supernova Cosmology Project} collaboration, R.~A. Knop et~al.,
  \emph{{New constraints on $\Omega_\mathrm{M}$, $\Omega_\Lambda$, and $w$ from
  an independent set of eleven high-redshift supernovae observed with HST}},
  \href{http://dx.doi.org/10.1086/378560}{\emph{Astrophys.~J.} {\bf 598} (2003)
  102}, [\href{http://arxiv.org/abs/astro-ph/0309368}{{\tt astro-ph/0309368}}].

\bibitem{eisensteinetal2005}
{\scshape SDSS} collaboration, D.~J. Eisenstein et~al., \emph{{Detection of the
  baryon acoustic peak in the large-scale correlation function of SDSS luminous
  red galaxies}}, \href{http://dx.doi.org/10.1086/466512}{\emph{Astrophys.~J.}
  {\bf 633} (2005) 560}, [\href{http://arxiv.org/abs/astro-ph/0501171}{{\tt
  astro-ph/0501171}}].

\bibitem{astieretal2006}
{\scshape SNLS} collaboration, P.~Astier et~al., \emph{{The Supernova legacy
  survey: Measurement of $\Omega_\mathrm{M}$, $\Omega_\Lambda$ and $w$ from the
  first year data set}},
  \href{http://dx.doi.org/10.1051/0004-6361:20054185}{\emph{Astron.~Astrophys.}
  {\bf 447} (2006) 31}, [\href{http://arxiv.org/abs/astro-ph/0510447}{{\tt
  astro-ph/0510447}}].

\bibitem{kowalskietal2008}
{\scshape Supernova Cosmology Project} collaboration, M.~Kowalski et~al.,
  \emph{{Improved Cosmological Constraints from New, Old and Combined Supernova
  Datasets}}, \href{http://dx.doi.org/10.1086/589937}{\emph{Astrophys.~J.} {\bf
  686} (2008) 749}, [\href{http://arxiv.org/abs/0804.4142}{{\tt 0804.4142}}].

\bibitem{wangwoodard2015b}
C.~L. Wang and R.~P. Woodard, \emph{{One-loop quantum electrodynamic correction
  to the gravitational potentials on de Sitter spacetime}},
  \href{http://dx.doi.org/10.1103/PhysRevD.92.084008}{\emph{Phys.~Rev.~D} {\bf
  92} (2015) 084008}, [\href{http://arxiv.org/abs/1508.01564}{{\tt
  1508.01564}}].

\bibitem{parkprokopecwoodard2016}
S.~Park, T.~Prokopec and R.~P. Woodard, \emph{{Quantum Scalar Corrections to
  the Gravitational Potentials on de Sitter Background}},
  \href{http://dx.doi.org/10.1007/JHEP01(2016)074}{\emph{JHEP} {\bf 01} (2016)
  074}, [\href{http://arxiv.org/abs/1510.03352}{{\tt 1510.03352}}].

\bibitem{froebverdaguer2016a}
M.~B. Fr{\"o}b and E.~Verdaguer, \emph{{Quantum corrections to the
  gravitational potentials of a point source due to conformal fields in de
  Sitter}}, \href{http://dx.doi.org/10.1088/1475-7516/2016/03/015}{\emph{JCAP}
  {\bf 1603} (2016) 015}, [\href{http://arxiv.org/abs/1601.03561}{{\tt
  1601.03561}}].

\bibitem{miaowoodard2006}
S.~P. Miao and R.~P. Woodard, \emph{{Gravitons Enhance Fermions during
  Inflation}},
  \href{http://dx.doi.org/10.1103/PhysRevD.74.024021}{\emph{Phys.~Rev.~D} {\bf
  74} (2006) 024021}, [\href{http://arxiv.org/abs/gr-qc/0603135}{{\tt
  gr-qc/0603135}}].

\bibitem{wangwoodard2015a}
C.~L. Wang and R.~P. Woodard, \emph{{Excitation of Photons by Inflationary
  Gravitons}},
  \href{http://dx.doi.org/10.1103/PhysRevD.91.124054}{\emph{Phys.~Rev.~D} {\bf
  91} (2015) 124054}, [\href{http://arxiv.org/abs/1408.1448}{{\tt 1408.1448}}].

\bibitem{tomboulis1977}
E.~Tomboulis, \emph{{$1/N$ expansion and renormalization in quantum gravity}},
  \href{http://dx.doi.org/10.1016/0370-2693(77)90678-5}{\emph{Phys.~Lett.~B}
  {\bf 70} (1977) 361}.

\bibitem{hartlehorowitz1981}
J.~B. Hartle and G.~T. Horowitz, \emph{{Ground-state expectation value of the
  metric in the $1/N$ or semiclassical approximation to quantum gravity}},
  \href{http://dx.doi.org/10.1103/PhysRevD.24.257}{\emph{Phys.~Rev.~D} {\bf 24}
  (1981) 257}.

\bibitem{hurouraverdaguer2004}
B.~L. Hu, A.~Roura and E.~Verdaguer, \emph{{Induced quantum metric fluctuations
  and the validity of semiclassical gravity}},
  \href{http://dx.doi.org/10.1103/PhysRevD.70.044002}{\emph{Phys.~Rev.~D} {\bf
  70} (2004) 044002}, [\href{http://arxiv.org/abs/gr-qc/0402029}{{\tt
  gr-qc/0402029}}].

\bibitem{osbornpetkou1994}
H.~Osborn and A.~C. Petkou, \emph{{Implications of conformal invariance in
  field theories for general dimensions}},
  \href{http://dx.doi.org/10.1006/aphy.1994.1045}{\emph{Annals~Phys.} {\bf 231}
  (1994) 311}, [\href{http://arxiv.org/abs/hep-th/9307010}{{\tt
  hep-th/9307010}}].

\bibitem{donoghueetal2002}
J.~F. Donoghue, B.~R. Holstein, B.~Garbrecht and T.~Konstandin, \emph{{Quantum
  corrections to the Reissner-Nordstr{\"o}m and Kerr-Newman metrics}},
  \href{http://dx.doi.org/10.1016/S0370-2693(02)01246-7}{\emph{Phys.~Lett.~B}
  {\bf 529} (2002) 132}, [\href{http://arxiv.org/abs/hep-th/0112237}{{\tt
  hep-th/0112237}}]. Erratum:
  \href{http://dx.doi.org/10.1016/j.physletb.2005.03.018}{\emph{Phys.~Lett.~B}
  \textbf{612} (2005) 311}.

\bibitem{witten2001}
E.~Witten, \emph{{Quantum gravity in de Sitter space}},  in \emph{{Strings
  2001: International Conference Mumbai, India, January 5-10, 2001}}, 2001.
\newblock \href{http://arxiv.org/abs/hep-th/0106109}{{\tt hep-th/0106109}}.

\bibitem{bousso2005}
R.~Bousso, \emph{{Cosmology and the S-matrix}},
  \href{http://dx.doi.org/10.1103/PhysRevD.71.064024}{\emph{Phys.~Rev.~D} {\bf
  71} (2005) 064024}, [\href{http://arxiv.org/abs/hep-th/0412197}{{\tt
  hep-th/0412197}}].

\bibitem{miaowoodard2012}
S.~P. Miao and R.~P. Woodard, \emph{{Issues Concerning Loop Corrections to the
  Primordial Power Spectra}},
  \href{http://dx.doi.org/10.1088/1475-7516/2012/07/008}{\emph{JCAP} {\bf 1207}
  (2012) 008}, [\href{http://arxiv.org/abs/1204.1784}{{\tt 1204.1784}}].

\bibitem{donnellygiddings2016}
W.~Donnelly and S.~B. Giddings, \emph{{Diffeomorphism-invariant observables and
  their nonlocal algebra}},
  \href{http://dx.doi.org/10.1103/PhysRevD.93.024030}{\emph{Phys.~Rev.~D} {\bf
  93} (2016) 024030}, [\href{http://arxiv.org/abs/1507.07921}{{\tt
  1507.07921}}]. Erratum:
  \href{http://dx.doi.org/10.1103/PhysRevD.94.029903}{\emph{Phys.~Rev.~D}
  \textbf{94} (2016) 029903}.

\bibitem{brunettietal2016}
R.~Brunetti, K.~Fredenhagen, T.-P. Hack, N.~Pinamonti and K.~Rejzner,
  \emph{{Cosmological perturbation theory and quantum gravity}},
  \href{http://dx.doi.org/10.1007/JHEP08(2016)032}{\emph{JHEP} {\bf 08} (2016)
  032}, [\href{http://arxiv.org/abs/1605.02573}{{\tt 1605.02573}}].

\bibitem{schwinger1961}
J.~S. Schwinger, \emph{{Brownian motion of a quantum oscillator}},
  \href{http://dx.doi.org/10.1063/1.1703727}{\emph{J.~Math.~Phys.} {\bf 2}
  (1961) 407}.

\bibitem{keldysh1965}
L.~V. Keldysh, \emph{{Diagram technique for nonequilibrium processes}},
  {\emph{Sov.~Phys.~JETP} {\bf 20} (1965) 1018}.
  [\href{http://www.jetp.ac.ru/cgi-bin/dn/e_020_04_1018.pdf}{Zh. Eksp. Teor.
  Fiz. \textbf{47} (1964) 1515}].

\bibitem{chousuhaoyu1985}
K.-c. Chou, Z.-b. Su, B.-l. Hao and L.~Yu, \emph{{Equilibrium and
  Nonequilibrium Formalisms Made Unified}},
  \href{http://dx.doi.org/10.1016/0370-1573(85)90136-X}{\emph{Phys.~Rept.} {\bf
  118} (1985) 1}.

\bibitem{higuchilee2009}
A.~Higuchi and Y.~C. Lee, \emph{{Conformally-coupled massive scalar field in de
  Sitter expanding universe with the mass term treated as a perturbation}},
  \href{http://dx.doi.org/10.1088/0264-9381/26/13/135019}{\emph{Class.~Quant.~Grav.}
  {\bf 26} (2009) 135019}, [\href{http://arxiv.org/abs/0903.3881}{{\tt
  0903.3881}}].

\bibitem{obukhovpuetzfeld2011}
Y.~N. Obukhov and D.~Puetzfeld, \emph{{Dynamics of test bodies with spin in de
  Sitter spacetime}},
  \href{http://dx.doi.org/10.1103/PhysRevD.83.044024}{\emph{Phys.~Rev.~D} {\bf
  83} (2011) 044024}, [\href{http://arxiv.org/abs/1010.1451}{{\tt 1010.1451}}].

\bibitem{camposverdaguer1994}
A.~Campos and E.~Verdaguer, \emph{{Semiclassical equations for weakly
  inhomogeneous cosmologies}},
  \href{http://dx.doi.org/10.1103/PhysRevD.49.1861}{\emph{Phys.~Rev.~D} {\bf
  49} (1994) 1861}, [\href{http://arxiv.org/abs/gr-qc/9307027}{{\tt
  gr-qc/9307027}}].

\bibitem{camposverdaguer1996}
A.~Campos and E.~Verdaguer, \emph{{Stochastic semiclassical equations for
  weakly inhomogeneous cosmologies}},
  \href{http://dx.doi.org/10.1103/PhysRevD.53.1927}{\emph{Phys.~Rev.~D} {\bf
  53} (1996) 1927}, [\href{http://arxiv.org/abs/gr-qc/9511078}{{\tt
  gr-qc/9511078}}].

\bibitem{fprv2013}
M.~B. Fr{\"o}b, D.~B. Papadopoulos, A.~Roura and E.~Verdaguer,
  \emph{{Nonperturbative semiclassical stability of de Sitter spacetime for
  small metric deviations}},
  \href{http://dx.doi.org/10.1103/PhysRevD.87.064019}{\emph{Phys.~Rev.~D} {\bf
  87} (2013) 064019}, [\href{http://arxiv.org/abs/1301.5261}{{\tt 1301.5261}}].

\bibitem{frv2014}
M.~B. Fr{\"o}b, A.~Roura and E.~Verdaguer, \emph{{Riemann correlator in de
  Sitter including loop corrections from conformal fields}},
  \href{http://dx.doi.org/10.1088/1475-7516/2014/07/048}{\emph{JCAP} {\bf 1407}
  (2014) 048}, [\href{http://arxiv.org/abs/1403.3335}{{\tt 1403.3335}}].

\bibitem{mtw}
C.~Misner, K.~Thorne and J.~A. Wheeler, \emph{{Gravitation}}.
\newblock \href{http://www.worldcat.org/search?q=isbn:9780716703440}{W.~H.~Freeman, San Francisco, 1973}.

\bibitem{mathisson1937}
M.~Mathisson, \emph{{Neue Mechanik materieller Systeme}},
  {\emph{Acta~Phys.~Polon.} {\bf 6} (1937) 163}. {English translation published
  in: \href{http://dx.doi.org/10.1007/s10714-010-0939-y}{Gen.~Rel.~Grav.
  \textbf{42} (2010) 1011}}.

\bibitem{papapetrou1951}
A.~Papapetrou, \emph{{Spinning test particles in general relativity. I.}},
  \href{http://dx.doi.org/10.1098/rspa.1951.0200}{\emph{Proc.~Roy.~Soc.~Lond.~A}
  {\bf 209} (1951) 248}.

\bibitem{tulczyjew1959}
W.~M. Tulczyjew, \emph{{Motion of multipole particles in general relativity
  theory}}, {\emph{Acta~Phys.~Pol.} {\bf 18} (1959) 393}.

\bibitem{tulczyjew1962}
B.~Tulczyjew and W.~M. Tulczyjew, \emph{On multipole formalism in general
  relativity}, in \emph{Recent Developments in General Relativity}, p.~465.
\newblock Pergamon Press, New York, USA, 1962.

\bibitem{taub1964}
A.~H. Taub, \emph{{Motion of test bodies in general relativity}},
  \href{http://dx.doi.org/10.1063/1.1704055}{\emph{J.~Math.~Phys.} {\bf 5}
  (1964) 112}.

\bibitem{dixon1964}
W.~G. Dixon, \emph{{A covariant multipole formalism for extended test bodies in
  general relativity}},
  \href{http://dx.doi.org/10.1007/BF02734579}{\emph{Nuovo~Cim.} {\bf 34} (1964)
  317}.

\bibitem{trautmann1965}
A.~Trautmann, \emph{{Lectures on general relativity}},  (Englewoods Cliffs,
  USA), Prentice Hall, 1965.
\newblock {Published for the Brandeis Univ. Summer Inst. Theoretical Physics
  1964, republished in:
  \href{http://dx.doi.org/10.1023/A:1015939926662}{Gen.~Rel.~Grav. \textbf{34}
  (2002) 721}}.

\bibitem{frenkel1926}
J.~Frenkel, \emph{{Die Elektrodynamik des rotierenden Elektrons}},
  \href{http://dx.doi.org/10.1007/BF01397099}{\emph{Z.~Phys.} {\bf 37} (1926)
  243}.

\bibitem{pirani1956}
F.~A.~E. Pirani, \emph{{On the physical significance of the Riemann tensor}},
  {\emph{Acta~Phys.~Polon.} {\bf 15} (1956) 389}. {Reprinted in:
  \href{http://dx.doi.org/10.1007/s10714-009-0787-9}{Gen.~Rel.~Grav.
  \textbf{41} (2009) 1215}}.

\bibitem{ohashi2003}
A.~Ohashi, \emph{{Multipole particle in relativity}},
  \href{http://dx.doi.org/10.1103/PhysRevD.68.044009}{\emph{Phys.~Rev.~D} {\bf
  68} (2003) 044009}, [\href{http://arxiv.org/abs/gr-qc/0306062}{{\tt
  gr-qc/0306062}}].

\bibitem{steinhoff2010}
J.~Steinhoff, \emph{{Canonical formulation of spin in general relativity}},
  \href{http://dx.doi.org/10.1002/andp.201000178}{\emph{Annalen~Phys.} {\bf
  523} (2011) 296}, [\href{http://arxiv.org/abs/1106.4203}{{\tt 1106.4203}}].

\bibitem{blanchet2011}
L.~Blanchet, \emph{{Post-Newtonian theory and the two-body problem}},
  \href{http://dx.doi.org/10.1007/978-90-481-3015-3_5}{\emph{Fundam.~Theor.~Phys.}
  {\bf 162} (2011) 125}, [\href{http://arxiv.org/abs/0907.3596}{{\tt
  0907.3596}}].

\bibitem{deriglazovramirez2015a}
A.~A. Deriglazov and W.~G. Ram{\'\i}rez, \emph{Lagrangian formulation for
  Mathisson-Papapetrou-Tulczyjew-Dixon (MPTD) equations},
  \href{http://dx.doi.org/10.1103/PhysRevD.92.124017}{\emph{Phys.~Rev.~D} {\bf
  92} (2015) 124017}, [\href{http://arxiv.org/abs/1509.04926}{{\tt
  1509.04926}}].

\bibitem{deriglazovramirez2015b}
A.~A. Deriglazov and W.~G. Ram{\'\i}rez, \emph{{Ultra-relativistic spinning
  particle and a rotating body in external fields}},
  \href{http://dx.doi.org/10.1155/2016/1376016}{\emph{Adv.~High~Energy~Phys.}
  {\bf 2016} (2016) 1376016}, [\href{http://arxiv.org/abs/1511.00645}{{\tt
  1511.00645}}].

\bibitem{abramobrandenbergermukhanov1997}
L.~R.~W. Abramo, R.~H. Brandenberger and V.~F. Mukhanov, \emph{{Energy-momentum
  tensor for cosmological perturbations}},
  \href{http://dx.doi.org/10.1103/PhysRevD.56.3248}{\emph{Phys.~Rev.~D} {\bf
  56} (1997) 3248}, [\href{http://arxiv.org/abs/gr-qc/9704037}{{\tt
  gr-qc/9704037}}].

\bibitem{nakamura2007}
K.~Nakamura, \emph{{Second-order gauge invariant cosmological perturbation
  theory: Einstein equations in terms of gauge invariant variables}},
  \href{http://dx.doi.org/10.1143/PTP.117.17}{\emph{Prog.~Theor.~Phys.} {\bf
  117} (2007) 17}, [\href{http://arxiv.org/abs/gr-qc/0605108}{{\tt
  gr-qc/0605108}}].

\bibitem{bardeen1980}
J.~M. Bardeen, \emph{{Gauge-invariant cosmological perturbations}},
  \href{http://dx.doi.org/10.1103/PhysRevD.22.1882}{\emph{Phys.~Rev.~D} {\bf
  22} (1980) 1882}.

\bibitem{duff1977}
M.~J. Duff, \emph{{Observations on Conformal Anomalies}},
  \href{http://dx.doi.org/10.1016/0550-3213(77)90410-2}{\emph{Nucl.~Phys.~B}
  {\bf 125} (1977) 334}.

\bibitem{toms1983}
D.~J. Toms, \emph{{The Effective Action and the Renormalization Group Equation
  in Curved Space-time}},
  \href{http://dx.doi.org/10.1016/0370-2693(83)90011-4}{\emph{Phys.~Lett.~B}
  {\bf 126} (1983) 37}.

\bibitem{kerr1963}
R.~P. Kerr, \emph{{Gravitational field of a spinning mass as an example of
  algebraically special metrics}},
  \href{http://dx.doi.org/10.1103/PhysRevLett.11.237}{\emph{Phys.~Rev.~Lett.}
  {\bf 11} (1963) 237}.

\bibitem{boyerlindquist1967}
R.~H. Boyer and R.~W. Lindquist, \emph{{Maximal analytic extension of the Kerr
  metric}}, \href{http://dx.doi.org/10.1063/1.1705193}{\emph{J.~Math.~Phys.}
  {\bf 8} (1967) 265}.

\bibitem{carter1973}
B.~Carter, \emph{{Black holes equilibrium states}},  in \emph{{Black Holes --
  Les astres occlus. Proceedings, École d'été de Physique Théorique Les
  Houches, France, August 1972}} (C.~deWitt and B.~deWitt, eds.), (New
  York/London/Paris), p.~61,
  \href{http://www.worldcat.org/search?q=isbn:9780677156101}{Gordon and
  Breach, 1973}.
\newblock Republished in:
  \href{http://dx.doi.org/10.1007/s10714-009-0888-5}{Gen.~Rel.~Grav.
  \textbf{41} (2009) 2873}.

\bibitem{gibbonshawking1977}
G.~W. Gibbons and S.~W. Hawking, \emph{{Cosmological Event Horizons,
  Thermodynamics, and Particle Creation}},
  \href{http://dx.doi.org/10.1103/PhysRevD.15.2738}{\emph{Phys.~Rev.~D} {\bf
  15} (1977) 2738}.

\bibitem{akcaymatzner2011}
S.~Akcay and R.~A. Matzner, \emph{{Kerr-de Sitter Universe}},
  \href{http://dx.doi.org/10.1088/0264-9381/28/8/085012}{\emph{Class.~Quant.~Grav.}
  {\bf 28} (2011) 085012}, [\href{http://arxiv.org/abs/1011.0479}{{\tt
  1011.0479}}].

\bibitem{gibbonsetal2005}
G.~W. Gibbons, H.~L{\"u}, D.~N. Page and C.~N. Pope, \emph{{The General Kerr-de
  Sitter metrics in all dimensions}},
  \href{http://dx.doi.org/10.1016/j.geomphys.2004.05.001}{\emph{J.~Geom.~Phys.}
  {\bf 53} (2005) 49}, [\href{http://arxiv.org/abs/hep-th/0404008}{{\tt
  hep-th/0404008}}].

\bibitem{froeb2016}
M.~B. Fr{\"o}b, \emph{{Quantum gravitational corrections for spinning
  particles}}, \href{http://dx.doi.org/10.1007/JHEP10(2016)051}{\emph{JHEP}
  {\bf 10} (2016) 051}, [\href{http://arxiv.org/abs/1607.03129}{{\tt
  1607.03129}}].

\bibitem{gazeaulachieze2006}
J.~P. Gazeau and M.~Lachi{\`e}ze-Rey, \emph{{Quantum field theory in de Sitter
  space: A Survey of recent approaches}}, {\emph{PoS} {\bf IC2006} (2006) 007},
  [\href{http://arxiv.org/abs/hep-th/0610296}{{\tt hep-th/0610296}}].

\bibitem{higuchi2008}
A.~Higuchi, \emph{{Decay of the free-theory vacuum of scalar field theory in de
  Sitter spacetime in the interaction picture}},
  \href{http://dx.doi.org/10.1088/0264-9381/26/7/072001}{\emph{Class.~Quant.~Grav.}
  {\bf 26} (2009) 072001}, [\href{http://arxiv.org/abs/0809.1255}{{\tt
  0809.1255}}].

\bibitem{brosepsteinmoschella2010}
J.~Bros, H.~Epstein and U.~Moschella, \emph{{Particle decays and stability on
  the de Sitter universe}},
  \href{http://dx.doi.org/10.1007/s00023-010-0042-7}{\emph{Ann.~H.~Poincar{\'e}}
  {\bf 11} (2010) 611}, [\href{http://arxiv.org/abs/0812.3513}{{\tt
  0812.3513}}].

\bibitem{higuchimarolfmorrison2011a}
A.~Higuchi, D.~Marolf and I.~A. Morrison, \emph{{Equivalence between Euclidean
  and in-in formalisms in de Sitter QFT}},
  \href{http://dx.doi.org/10.1103/PhysRevD.83.084029}{\emph{Phys.~Rev.~D} {\bf
  83} (2011) 084029}, [\href{http://arxiv.org/abs/1012.3415}{{\tt 1012.3415}}].

\bibitem{marolfmorrison2011a}
D.~Marolf and I.~A. Morrison, \emph{{The IR stability of de Sitter QFT: results
  at all orders}},
  \href{http://dx.doi.org/10.1103/PhysRevD.84.044040}{\emph{Phys.~Rev.~D} {\bf
  84} (2011) 044040}, [\href{http://arxiv.org/abs/1010.5327}{{\tt 1010.5327}}].

\bibitem{marolfmorrison2011b}
D.~Marolf and I.~A. Morrison, \emph{{The IR stability of de Sitter QFT:
  Physical initial conditions}},
  \href{http://dx.doi.org/10.1007/s10714-011-1233-3}{\emph{Gen.~Rel.~Grav.}
  {\bf 43} (2011) 3497}, [\href{http://arxiv.org/abs/1104.4343}{{\tt
  1104.4343}}].

\bibitem{hollands2013}
S.~Hollands, \emph{{Correlators, Feynman diagrams, and quantum no-hair in
  deSitter spacetime}},
  \href{http://dx.doi.org/10.1007/s00220-012-1653-2}{\emph{Commun.~Math.~Phys.}
  {\bf 319} (2013) 1}, [\href{http://arxiv.org/abs/1010.5367}{{\tt
  1010.5367}}].

\bibitem{hollands2012}
S.~Hollands, \emph{{Massless interacting quantum fields in deSitter
  spacetime}},
  \href{http://dx.doi.org/10.1007/s00023-011-0140-1}{\emph{Ann.~H.~Poincar{\'e}}
  {\bf 13} (2012) 1039}, [\href{http://arxiv.org/abs/1105.1996}{{\tt
  1105.1996}}].

\bibitem{marolfmorrisonsrednicki2013}
D.~Marolf, I.~A. Morrison and M.~Srednicki, \emph{{Perturbative S-matrix for
  massive scalar fields in global de Sitter space}},
  \href{http://dx.doi.org/10.1088/0264-9381/30/15/155023}{\emph{Class.~Quant.~Grav.}
  {\bf 30} (2013) 155023}, [\href{http://arxiv.org/abs/1209.6039}{{\tt
  1209.6039}}].

\bibitem{koraitanaka2013}
Y.~Korai and T.~Tanaka, \emph{{Quantum field theory in the flat chart of de
  Sitter space}},
  \href{http://dx.doi.org/10.1103/PhysRevD.87.024013}{\emph{Phys.~Rev.~D} {\bf
  87} (2013) 024013}, [\href{http://arxiv.org/abs/1210.6544}{{\tt 1210.6544}}].

\bibitem{moschella2016}
U.~Moschella, \emph{{Infrared surprises in the de Sitter universe}},
  \href{http://dx.doi.org/10.1142/S0218271816410200}{\emph{Int.~J.~Mod.~Phys.}
  {\bf D25} (2016) 1641020}, [\href{http://arxiv.org/abs/1210.4815}{{\tt
  1210.4815}}].

\bibitem{belokognefolacciqueva2016}
A.~Belokogne, A.~Folacci and J.~Queva, \emph{{Stueckelberg massive
  electromagnetism in de Sitter and anti–de Sitter spacetimes: Two-point
  functions and renormalized stress-energy tensors}},
  \href{http://dx.doi.org/10.1103/PhysRevD.94.105028}{\emph{Phys.~Rev.~D} {\bf
  94} (2016) 105028}, [\href{http://arxiv.org/abs/1610.00244}{{\tt
  1610.00244}}].

\bibitem{kahyawoodard2007}
E.~O. Kahya and R.~P. Woodard, \emph{{Quantum Gravity Corrections to the One
  Loop Scalar Self-Mass during Inflation}},
  \href{http://dx.doi.org/10.1103/PhysRevD.76.124005}{\emph{Phys.~Rev.~D} {\bf
  76} (2007) 124005}, [\href{http://arxiv.org/abs/0709.0536}{{\tt 0709.0536}}].

\bibitem{higuchimarolfmorrison2011b}
A.~Higuchi, D.~Marolf and I.~A. Morrison, \emph{{de Sitter invariance of the dS
  graviton vacuum}},
  \href{http://dx.doi.org/10.1088/0264-9381/28/24/245012}{\emph{Class.~Quant.~Grav.}
  {\bf 28} (2011) 245012}, [\href{http://arxiv.org/abs/1107.2712}{{\tt
  1107.2712}}].

\bibitem{moratsamiswoodard2012}
P.~J. Mora, N.~C. Tsamis and R.~P. Woodard, \emph{{Weyl-Weyl Correlator in de
  Donder Gauge on de Sitter}},
  \href{http://dx.doi.org/10.1103/PhysRevD.86.084016}{\emph{Phys.~Rev.~D} {\bf
  86} (2012) 084016}, [\href{http://arxiv.org/abs/1205.4466}{{\tt 1205.4466}}].

\bibitem{frv2012}
M.~B. Fr{\"o}b, A.~Roura and E.~Verdaguer, \emph{{One-loop gravitational wave
  spectrum in de Sitter spacetime}},
  \href{http://dx.doi.org/10.1088/1475-7516/2012/08/009}{\emph{JCAP} {\bf 1208}
  (2012) 009}, [\href{http://arxiv.org/abs/1205.3097}{{\tt 1205.3097}}].

\bibitem{battistaesposito2014}
E.~Battista and G.~Esposito, \emph{{Restricted three-body problem in
  effective-field-theory models of gravity}},
  \href{http://dx.doi.org/10.1103/PhysRevD.89.084030}{\emph{Phys.~Rev.~D} {\bf
  89} (2014) 084030}, [\href{http://arxiv.org/abs/1402.2931}{{\tt 1402.2931}}].

\bibitem{battistaetal2015}
E.~Battista, S.~Dell'Agnello, G.~Esposito and J.~Simo, \emph{{Quantum effects
  on Lagrangian points and displaced periodic orbits in the Earth-Moon
  system}},
  \href{http://dx.doi.org/10.1103/PhysRevD.91.084041}{\emph{Phys.~Rev.~D} {\bf
  91} (2015) 084041}, [\href{http://arxiv.org/abs/1501.02723}{{\tt
  1501.02723}}]. Erratum:
  \href{http://dx.doi.org/10.1103/PhysRevD.93.049902}{\emph{Phys.~Rev.~D}
  \textbf{93} (2016) 049902}.

\bibitem{battistaetal2017}
E.~Battista, A.~Tartaglia, G.~Esposito, D.~Lucchesi, M.~L. Ruggiero, P.~Valko
  et~al., \emph{{Quantum time delay in the gravitational field of a rotating
  mass}},  \href{http://arxiv.org/abs/1703.08095}{{\tt 1703.08095}}.

\bibitem{weinberg2005}
S.~Weinberg, \emph{{Quantum contributions to cosmological correlations}},
  \href{http://dx.doi.org/10.1103/PhysRevD.72.043514}{\emph{Phys.~Rev.~D} {\bf
  72} (2005) 043514}, [\href{http://arxiv.org/abs/hep-th/0506236}{{\tt
  hep-th/0506236}}].

\bibitem{starobinsky1994}
A.~A. Starobinsky and J.~Yokoyama, \emph{{Equilibrium state of a
  selfinteracting scalar field in the de Sitter background}},
  \href{http://dx.doi.org/10.1103/PhysRevD.50.6357}{\emph{Phys.~Rev.~D} {\bf
  50} (1994) 6357--6368}, [\href{http://arxiv.org/abs/astro-ph/9407016}{{\tt
  astro-ph/9407016}}].

\bibitem{woodard2005}
R.~P. Woodard, \emph{{A Leading logarithm approximation for inflationary
  quantum field theory}},
  \href{http://dx.doi.org/10.1016/j.nuclphysbps.2005.04.056}{\emph{Nucl.~Phys.~Proc.~Suppl.}
  {\bf 148} (2005) 108}, [\href{http://arxiv.org/abs/astro-ph/0502556}{{\tt
  astro-ph/0502556}}].

\bibitem{riottosloth2008}
A.~Riotto and M.~S. Sloth, \emph{{On Resumming Inflationary Perturbations
  beyond One-loop}},
  \href{http://dx.doi.org/10.1088/1475-7516/2008/04/030}{\emph{JCAP} {\bf 0804}
  (2008) 030}, [\href{http://arxiv.org/abs/0801.1845}{{\tt 0801.1845}}].

\bibitem{rajaraman2010}
A.~Rajaraman, \emph{{On the proper treatment of massless fields in Euclidean de
  Sitter space}},
  \href{http://dx.doi.org/10.1103/PhysRevD.82.123522}{\emph{Phys.~Rev.~D} {\bf
  82} (2010) 123522}, [\href{http://arxiv.org/abs/1008.1271}{{\tt 1008.1271}}].

\bibitem{benekemoch2013}
M.~Beneke and P.~Moch, \emph{{On ``dynamical mass'' generation in Euclidean de
  Sitter space}},
  \href{http://dx.doi.org/10.1103/PhysRevD.87.064018}{\emph{Phys. Rev.} {\bf
  D87} (2013) 064018}, [\href{http://arxiv.org/abs/1212.3058}{{\tt
  1212.3058}}].

\bibitem{serreauparentani2013}
J.~Serreau and R.~Parentani, \emph{{Nonperturbative resummation of de Sitter
  infrared logarithms in the large-$N$ limit}},
  \href{http://dx.doi.org/10.1103/PhysRevD.87.085012}{\emph{Phys.~Rev.~D} {\bf
  87} (2013) 085012}, [\href{http://arxiv.org/abs/1302.3262}{{\tt 1302.3262}}].

\bibitem{gautierserreau2013}
F.~Gautier and J.~Serreau, \emph{{Infrared dynamics in de Sitter space from
  Schwinger-Dyson equations}},
  \href{http://dx.doi.org/10.1016/j.physletb.2013.10.072}{\emph{Phys.~Lett.~B}
  {\bf 727} (2013) 541}, [\href{http://arxiv.org/abs/1305.5705}{{\tt
  1305.5705}}].

\bibitem{youssefkreimer2014}
A.~Youssef and D.~Kreimer, \emph{{Resummation of infrared logarithms in de
  Sitter space via Dyson-Schwinger equations: the ladder-rainbow
  approximation}},
  \href{http://dx.doi.org/10.1103/PhysRevD.89.124021}{\emph{Phys.~Rev.~D} {\bf
  89} (2014) 124021}, [\href{http://arxiv.org/abs/1301.3205}{{\tt 1301.3205}}].

\bibitem{lopeznaciretal2016}
D.~López~Nacir, F.~D. Mazzitelli and L.~G. Trombetta, \emph{{O(N) model in
  Euclidean de~Sitter space: beyond the leading infrared approximation}},
  \href{http://dx.doi.org/10.1007/JHEP09(2016)117}{\emph{JHEP} {\bf 09} (2016)
  117}, [\href{http://arxiv.org/abs/1606.03481}{{\tt 1606.03481}}].

\bibitem{collinsholman2005}
H.~Collins and R.~Holman, \emph{{Renormalization of initial conditions and the
  trans-Planckian problem of inflation}},
  \href{http://dx.doi.org/10.1103/PhysRevD.71.085009}{\emph{Phys.~Rev.~D} {\bf
  71} (2005) 085009}, [\href{http://arxiv.org/abs/hep-th/0501158}{{\tt
  hep-th/0501158}}].

\bibitem{collins2013}
H.~Collins, \emph{{Initial state propagators}},
  \href{http://dx.doi.org/10.1007/JHEP11(2013)077}{\emph{JHEP} {\bf 11} (2013)
  077}, [\href{http://arxiv.org/abs/1309.2656}{{\tt 1309.2656}}].

\bibitem{collinsholmanverdanyan2014}
H.~Collins, R.~Holman and T.~Vardanyan, \emph{{Renormalizing an initial
  state}}, \href{http://dx.doi.org/10.1007/JHEP10(2014)124}{\emph{JHEP} {\bf
  10} (2014) 124}, [\href{http://arxiv.org/abs/1408.4801}{{\tt 1408.4801}}].

\bibitem{peskinschroeder}
M.~E. Peskin and D.~V. Schroeder, \emph{{An Introduction To Quantum Field
  Theory}}.
\newblock \href{http://www.worldcat.org/search?q=isbn:9780201503975}{Addison-Wesley, Reading, MA, 1995}.

\end{thebibliography}\endgroup

\end{document}